\documentclass[a4paper,fleqn,usenatbib]{mnras}

\usepackage{amssymb}
\usepackage{amsmath}
\usepackage{graphicx}
\usepackage{natbib}
\usepackage{hyperref}
\usepackage{pdflscape}
\usepackage{longtable}
\usepackage{multicol}
\usepackage{multirow}
\usepackage{url}
\usepackage{graphicx}
\usepackage{color}
\usepackage{txfonts}

\DeclareRobustCommand{\ION}[2]{%
\relax\ifmmode
\ifx\testbx\f@series
{\mathbf{#1\,\mathsc{#2}}}\else
{\mathrm{#1\,\mathsc{#2}}}\fi
\else\textup{#1\,{\mdseries\textsc{#2}}}%
\fi}
\newcommand{\lam}{$\lambda$}

\newcommand{\nii}{[\ION{N}{ii}]}

\newcommand{\oiii}{[\ION{O}{iii}]}

\newcommand{\Ha}{$\rm{H}\alpha$}
\newcommand{\Hb}{$\rm{H}\beta$}

\newcommand{\HII}{\ion{H}{ii}}

\newcommand{\Ssfr}{$\Sigma_{\rm SFR}$\,}
\newcommand{\Sst}{$\Sigma_{\rm *}$\,}
\newcommand{\Sgas}{$\Sigma_{\rm mol}$\,}
\newcommand{\Srgas}{$\Sigma_{\rm gas}$\,}

\newcommand{\eSsfr}{\Sigma_{\rm SFR}}
\newcommand{\eSst}{\Sigma_{\rm *}}
\newcommand{\eSgas}{\Sigma_{\rm mol}}

\title[]{The EDGE-CALIFA survey: The local and global relations between \Sst, \Ssfr and \Sgas that regulate star-formation}

\author[S.F.S\'anchez et al.]{S.F. S\'anchez$^{1}$, 
J.K. Barrera-Ballesteros$^{1}$, 
D. Colombo$^{2}$
T. Wong$^{3}$,
A. Bolatto$^{4}$,\newauthor
E. Rosolowsky$^{5}$,
S. Vogel$^{4}$,
R. Levy$^{4}$,
V. Kalinova$^2$,
P. Alvarez-Hurtado$^{1}$,
Y. Luo$^{3}$,
Y. Cao$^{3}$
\\
$^1$Instituto de Astronom\'ia, Universidad Nacional Aut\'onoma de  M\'exico, A.~P. 70-264, C.P. 04510, M\'exico, D.F., Mexico \\
$^2$ Max-Planck-Institut f\"ur Radioastronomie, Auf dem Hügel 69, 53121 Bonn, Germany\\
$^3$ Department of Astronomy, University of Illinois, Urbana, IL 61801, USA \\
$^4$ Department of Astronomy, University of Maryland, College Park, MD 20742, USA\\
$^5$ Department of Physics, University of Alberta, 4-183 CCIS, Edmonton, Alberta, Canada\\
}

\date{Accepted XXX. Received YYY; in original form ZZZ}

\pubyear{2020}

\begin{document}
\label{firstpage}
\pagerange{\pageref{firstpage}--\pageref{lastpage}}
\maketitle

\begin{abstract}

  We present a new characterization of the relations between star-formation rate, stellar mass and
  molecular gas mass surface densities at different spatial scales across galaxies (from
  galaxy wide to kpc-scales). To do so we make use of the largest
  sample combining spatially-resolved spectroscopic information with CO observations,
  provided by the EDGE-CALIFA survey, together with new single dish CO
  observations obtained by APEX. We show that those relations are
  the same at the different explored scales, sharing the same
  distributions for the explored data, with similar
  slope, intercept and scatter (when characterized by a simple
  power-law). From this analysis, we propose that these relations are the projection of a single relation between the three properties that
  follows a distribution well described by a line in the three-dimension
  parameter space. Finally, we show that observed secondary relations between the residuals and the considered parameters are fully explained by the correlation between the uncertainties, and therefore have no physical origin.  
  We discuss these results in the context of the  hypothesis of self-regulation of the star-formation process.

\end{abstract}


\section{Introduction}
\label{sec:intro}

Star-formation is likely the central physical process
defining the nature and evolution of galaxies.  
Indeed, gas clouds trapped in dark-matter halos not
forming stars are considered as failed galaxies
\citep[e.g.][]{whit92}. This process requires very particular physical
conditions that allow first, the formation of molecular clouds from the
more diffuse atomic gas content (mostly HI), and second the fragmentation of those
clouds that collapse in order to reach the densities (and temperatures)
required to ignite the thermonuclear reactions that define stars.
Star-forming galaxies, i.e., those that actively form stars across
their optical extent, follow three well known scaling relations
between their global star-formation rate, gas mass, and stellar mass
content, whose nature is still not fully understood.

The relation between the star-formation rate (SFR) and the { cold}
gas mass ({ atomic and molecular}) is known as the
Schmidt-Kennicutt law (SK-law, or simply, the star-formation-law or
SF-law). { It was first proposed by \citet{schmidt59} as a relation between the SFR
  and the mass of cold gas in a certain volume, with the form of a
  simple power-law, based on purely theoretical considerations.  He
  estimates the slope of this power-law between 2-3, based on the
  comparison between the dispersion in the vertical direction of our
  galaxy of both the neutral hydrogen and the number of young stars.
  The SK-law } was empirically confirmed as a relation between the
average SFR and { cold} gas surface densities, { atomic plus
  molecular}, averaged over the optical extent of galaxies. Thus, this
is a relation between two {\it intensive} quantities (\Ssfr and
\Srgas), i.e., quantities that do not depend on the size of
galaxies. { A initial slope of 1.3$\pm$0.3 was found by
  \citet{kennicutt89a}, with values between 0.9-1.7 reported in
  subsequent studies using different estimations of the SFR and the
  cold gas content \citep[as reviewed in][]{kennicutt98}.}A simple
free-fall timescale argument was proposed by this author to explain
this relation, and the value of its exponent. { Based on a
  compilation of literature data \citet{kennicutt98} proposed a relation with slope of
  1.4$\pm$0.15, and a dispersion of $\sim$0.28 dex, in \Ssfr, using the total cold gas density (i.e., \Srgas). A similar relation} has been found to hold not only galaxy wide,
but also at local/resolved scales (down to $\sim$500 pc), {
  but only when using} the molecular gas surface density { (i.e., \Sgas)}. {
  This relation, known as resolved Schmidt-Kennicutt law or rSK-law, presents
  a slope slightly lower or near to $\sim$1 and a dispersion of the order
  of $\sim$0.2 dex in \Ssfr \citep[e.g][]{wong02,Kennicutt07,bigiel08,leroy13},
  as recently reviewed in \citet{ARAA}.}

Large galaxy spectroscopic and imaging surveys like the Sloan Digital
Sky Survey \citep[SDSS,][]{york+2000}, reveal a tight relation between
the integrated SFR and stellar mass in galaxies, known as the
star-formation main sequence \citep[SFMS,
e.g.][]{brin04,renzini15}. Like the SK-law, this relation follows a
power-law, with a power/slope near or slightly lower than $\sim$1 at
z$\sim$0. However, contrary to the SK-law, it is well documented that
the SFMS evolves with time \citep[e.g.][]{speagle14,rodriguez16}, as
galaxies become less massive (in stellar mass) and exhibit larger SFR
following the cosmological increase in SFR density
\citep[e.g.][]{madau14,sanchez18b}. { The zero-point of the
  relation is essentially constant in the nearby Universe ($z<$0.15),
  followed by a strong increase from $z\sim$0.2 to $z\sim$2. The slope
  presents a similar trend, but with a weaker or no evolution,
  depending on the authors \citep[e.g, Fig. 7 and 8 from
  ][]{sanchez18b}.} In contrast with the SK-law, the SFMS was first
expressed as a relation between {\it extensive} quantities (integrated
M$_*$ and SFR), i.e., between parameters which value depends on the
size of galaxies. \citet{sanchez13} and \citet{wuyts13} first reported
a similar relation between the SFR and M$_*$ surface densities of
star-forming regions within galaxies found at kpc-scales (i.e., \Ssfr
vs. \Sst). This relation between intensive quantities, known as the
resolved SFMS (or rSMFS), was explored in detail by \citet{mariana16},
using a sample of galaxies with spatial resolved spectroscopic
information extracted from the { Calar Alto Legacy Integral Field
  Area } (CALIFA) integral-field spectroscopy (IFS) survey
\citep{sanchez12}. They found that this relation presents a similar
shape at a kpc-scale as the global one and a similar scatter
($\sim$0.25 dex). A possible morphological dependence of the shape of
{ the global and local/resolved SFMS relations} has been reported
in different studies \citep[e.g.][]{rosa16,catalan17,mariana19}. More
recently, the decomposition between the disk and bulge components in
galaxies has shown that { the reported differences by morphological
  types} may be artificial: { \citet{jairo19} showed that the SFMS
  relations for the disk component of both late- and early-type
  galaxies are statistically indistinguishable, despite the fact that
  a clear difference is appreciated if the integrated quantities
  (disk+bulge) are adopted. In summary, the disks of galaxies with
  different morphologies present the same SFMS relations. }

Finally, a similar relationship has been described between the
molecular gas and stellar masses in SF galaxies
\citep[e.g.][]{saint16,calette18}. This relation, known now as the
Molecular Gas Main Sequence (MGMS), has not been explored as much as
the previous two. In its spatially resolved form it was only recently
described as a power-law relation between the surface
densities of both quantities by \citet{lin19} (e.g. \Sgas
vs. \Sst). In that study, { the authors} used a limited sample of just 14
galaxies extracted from the MaNGA \citep{bundy15} IFS survey, observed
with the Atacama Large Millimeter Array (ALMA) to provide kpc-scale
CO-mapping \citep[as part of the ALMaQUEST compilation][]{lin20}. {
  They found that } this relation exhibits a slope $\sim$1 and a scatter of
$\sim$0.2 dex (the so-called rMGMS relation). { A more recent
  exploration by \citet{ellison21a}, increasing the number of studied
  ALMaQUEST galaxies by a factor two, provided a similar result.}

The connection between the global and local/resolved versions of the
three relations has been a topic of study in different
studies. \citet{bolatto17} first showed that a simple parametrization
describing the global intensive SK-law follows the observed trend of
the local/resolved distribution of the two considered parameters.  In
the case of the SFMS, the correspondence between the global and
local/resolved relations was first explored by \citet{pan18} and
\citet{mariana19}. More recently, \citet{sanchez20}, using a large
compilation of galaxies observed using IFS at kpc-scales ($\sim$8,000
galaxies), and an indirect proxy for the molecular gas, demonstrated
that the global and local/resolved versions of the MGMS follow similar
distributions, when the global relation is presented in its intensive
form \citep[i.e., the average surface density across the optical
extent of galaxies, following][]{mariana19}.

The nature of the scatter described in the three relations and the
existence of possible secondary relations with other parameters, that
may drive this scatter, is an important topic of study. Whether the
morphology, the gas content, the star-formation efficiency (SFE) {,
  or any other property still not explored in the literature} modulate
the relation between the three parameters is of a key importance to
understand the processes that regulate SF in galaxies (and regions
within galaxies). As indicated before, early results suggest that the
SFMS may depend on the morphology \citep[][]{catalan17}.{ However,
  the results were not totally conclusive. Some hints of dependence
  was reported for the rSFMS \citep[e.g.][]{mariana19} with the
  morphology, while clear variations were reported by more recent
  explorations \citep[]{ellison21a}. Finally, it has not been explored
  if these differences in the rSFMS are induced by the presence of the bulge,
  like it was found by \citet{jairo19} for the integrated/global SFMS relation (as quoted before). On the other hand, it is known that as star-formation
  halts/quenches in galaxies (and regions within galaxies), they
  depart from the explored relations, with the gas fraction
  ($f_{gas}$) being the major driver of this separation
  \citep[e.g.][]{saint16,calette18,colombo18,sanchez18,sanchez20,colombo20}. }
{ Although the separation from the three relations traced by retired galaxies
is directly related to a lack of (molecular) gas,}  the
dispersion within the rSFMS { (i.e., among SFGs)} was reported to depend both on the SFE and
the gas fraction \citep{ellison20,colombo20,ellison21a}. The relative importance
of both parameters is still unclear in driving this dispersion,
although the SFE seems to out-rank the gas fraction\citep[][,
Fig. 4]{ellison20}. { The presence of additional parameters that
modulate these relations may indicate the existence of a generalized
star-formation law from which the observed ones are just projections in a more
limited space of parameters \citep[e.g.][]{shi18,dey19,jkbb21a}. On the other hand, the need of additional parameters may indicate the
existence of a so-far hidden parameter that regulates the three
relations \citep[e.g., the gas pressure][]{jkbb21b}}

In this article we attempt to characterize the three relations in
their global intensive and local/resolved forms using a large and
statistically significant sample of galaxies observed using both IFS and
resolved and aperture limited molecular gas mapping. To do so we
use the extended CALIFA sample \citep{sanchez12}, in combination with
the recent single dish CO-mapping provided by APEX \citep{colombo20}
and the spatially resolved CO-mapping provided by the EDGE survey
\citep{bolatto17}. { This compilation comprises the largest available dataset
of the three considered parameters, including estimations covering the entire galaxies (CALIFA),
aperture limited (APEX), and spatially resolved (at kpc-scales, EDGE).
The three datasets have different spatial resolutions and they cover different
regimes of the same galaxies (or a subset of them). We will treat them
separately to narrow-down the effects of resolution and apertures in the
explored relations, comparing the results derived from each of them when needed}.
In addition
we explore the correspondence between the global intensive and
local/resolved relations. Particular care has been taken in the
interpretation of possible secondary relations and the effects of
errors, which has not been addressed in detail before.
The structure of this articles is as follows: the galaxy samples
and adopted datasets are described in Sec. \ref{sec:sample}, with
details of the optically derived parameters included in
Sec. \ref{sec:opt} and the CO-derived ones in Sec. \ref{sec:co}; a
summary of the main properties of the different galaxy subsamples,
compared with the ones adopted by previous explorations, is presented
in Sec. \ref{sec:main}; the analysis performed on the data is
described in Sec. \ref{sec:ana}, with a description of the possible
secondary relations included in Sec. \ref{sec:sec}; the effects of errors
in the apparent generation of these relations are described in Sec. \ref{sec:real}. Finally, we present the main conclusions and a discussion of the results in Sec. \ref{sec:dis}.

Throughout this article we assume the standard $\Lambda$ Cold 
Dark Matter cosmology with the parameters: $H_0$=71 km/s/Mpc, $\Omega_M$=0.27, 
$\Omega_\Lambda$=0.73.




\section{Sample and data}
\label{sec:sample}

The galaxies explored in this study were extracted from the
extended CALIFA sample \citep[eCALIFA,][]{espi20,lacerda20}. The Calar
Alto Legacy Integral Field Area survey \citep{sanchez12} is a survey
of galaxies in the nearby universe ($z<0.1$), observed at the 3.5m
telescope of the Calar Alto observatory, using the PPAK \citep{kelz06}
wide Integral Field Unit of the PMAS \citep{roth05} spectrograph. PPAK
offered one of the largest Field-of-views of currently existing IFUs
(74$\arcsec$$\times$64$\arcsec$), sampled with a
$\sim$60\% covering factor by
2.7$\arcsec$ fibers. Adopting a three dithering observing scheme { (i.e., three pointings
  with an offset smaller than the distance between fibers)} the
complete Field-of-View (FoV) is sampled, providing a final point-spread-function (PSF)
full-width at half-maximum (FWHM) of
$\sim$2.5$\arcsec$ \citep{dr2}.  Two
observational setups were adopted: the intermediate resolution V1200
setup (3700-4800\AA,
$\lambda/\Delta\lambda\sim$1650) and the low resolution V500 setup
(3745-7500 \AA,
$\lambda/\Delta\lambda\sim$850); the latter in the one used in this
study.

Galaxies were diameter-selected to match their optical extent with the
FoV of the IFU, covering up to 2.5 effective radius in all the
galaxies \citep{walcher14}. Additional cuts were imposed on the
original mother sample to exclude dwarf galaxies and extremely massive
ones. { As explained in \citet{sanchez12} and \citet{walcher14} in
  detail, dwarfs
  (M$<$10$^7$M$_\odot$) were originally excluded since their evolution
  is known to be different than more massive galaxies
  (M$>$10$^{8.5}$M$_\odot$), and they would dominate the number of
  objects if a pure diameter selection is performed (as they are far
  more numerous). On the other hand, extremelly massive galaxies are
  very rare in number in the nearby universe and for the covered
  volume no representative sample could be selected impossing a
  diameter selection. The cuts guarantee the statistical significance
  of the CALIFA mother sample.}

At the completion of the CALIFA survey, a set of subsamples were
observed to increase the number of certain galaxy types either
excluded due to the cuts { indicated before (dwarfs, large ellipticals)}
or resulting in low numbers in the final sample, or
additional galaxies of particular interest (like hosts of recent
supernovae). These extended samples fulfill the main selection
criteria of the original sample (low-redshift, diameter selection),
increasing the number statistics. All of them were observed using the
same instrumental setup, with the same observing strategy and reduced
and analyzed in a similar way. A particularly large number of galaxies
in these extended samples ($\sim$100) correspond to ongoing SN host
exploration by the { PMAS/PPak Integral-field Supernova Hosts Compilation} \citep[PISCO][]{pisco}. { As discussed in
  \citet{sanchez16}, all these galaxies where included in the CALIFA database,
as a set of extended subsamples.}

{ All together, the final extended CALIFA sample
  \citep{espi20,lacerda20}} comprises, so far, 941 galaxies with good
quality observations using the V500 setup\footnote{the current list of
  analyzed galaxies plus the CALIFA pilot sample can be consulted in
  the webpage:
  \url{http://ifs.astroscu.unam.mx/CALIFA/V500/list_Pipe3D_clean.php}}.
All data were reduced using version 2.2 of the CALIFA pipeline
\citep{sanchez16}, that performs all the usual reduction steps for
fiber-fed integral field spectroscopy \citep{sanchez06a}, including
fiber tracing, extraction, wavelength calibration, fiber-to-fiber
corrections, flux calibration and spatial registration. The final
product of the data-reduction is a cube with the spatial information
registered in the {\it x} and {\it y} axis, and the spectral one in
the {\it z} axis . The final reconstructed datacube has a sampling of
1$\arcsec$ per spaxel, large enough to correctly sample the final PSF
{ \citep[according to the Nyquist–Shannon sampling theorem,][]{nyquist28,shannon1948}}, without a large oversampling (and the corresponding increase of the
co-variance). The final datacubes have an astrometry accuracy of
$\sim$0.5$\arcsec$, an absolute photometric calibration precision of
$\sim$8\% ({ and a $\sim$5\% relative from blue to red, what quantifies the color effect in the spectra}), and a depth of
$r\sim$23.6 mag/arcsec$^2$. Details on the data reduction and the
quality of the data can be found in
\citet{sanchez12,husemann13,rgb15,sanchez16}.

\subsection{Optical derived parameters}
\label{sec:opt}

The IFS data provided by the CALIFA observations are analyzed using
the {\sc Pipe3D} dedicated pipeline \citep{pipe3d,pipe3d_ii}, to
extract the most relevant physical parameters of the stellar
population { (luminosity weighted ages and metallicities, dust attenuation, light distribution by ages, stellar velocity and velocity dispersion...)} and emission lines { (line fluxes, equivalent widths, ionized gas velocity and velocity dispersion...)} .  Details of this pipeline, explanations 
regarding the derivations of the different parameters, reliability tests,
and examples of its use have been extensively given in many different
articles \citep[e.g.][for citing just a
few]{mariana16,ibarra19,bluck19,ellison20,lacerda20,ARAA}.  In summary,
the pipeline
performs an analysis of the stellar population and emission line
properties by modelling each spectrum in the cube with a combination
of a set of single-stellar populations models (SSPs), using a
particular stellar library, and a set of Gaussian functions to
describe the emission lines. Stellar population models are convolved
with a line-of-sight velocity distribution to recover the stellar
kinematics, while the gas kinematics are recovered by the Gaussian
fitting itself. The stellar dust attenuation is derived in an iterative
procedure, using the same optical spectra. The procedure involves spatially binning the data
to increase the signal-to-noise above a threshold { of $\sim$50 at which the
stellar population analysis provides with reliable results based on simulations \citep{pipe3d_ii}}, and a dezonification to
recover a model for the original sampled spaxels
\citep{cidfernandes:2013}. In addition to this modelling, the stellar-subtracted datacube (the so-called gas-pure cube) is analyzed based on
a moment analysis to recover  
the properties of the
emission lines in more detail (i.e., flux intensity, equivalent width, velocity and velocity dispersion). The {\sc Pipe3D} pipeline was developed to analyse IFS
data of different origin, and has been tested with CALIFA
\citep[e.g.][]{mariana16,sanchez17a}, MaNGA
\citep[e.g.][]{ibarra16,mariana19,lin19}, SAMI \citep[e.g.][]{sanchez19} and
MUSE data \citep[e.g.][]{carlos20}.

The most relevant parameters for the current study derived based on the {\sc Pipe3D} analysis are: (i) the stellar mass; (ii) the star-formation rate
and (iii) a proxy for the molecular gas { that will be explained in detail below}. All these three quantities are derived
both integrated across the FoV of the IFS data and spatially resolved, spaxel-to-spaxel,
as surface densities. We summarise here how these quantities are derived.
For a more detailed explanation, we refer the reader to \citet{sanchez20}, in which
the derivation is explained in more detail:

{ Stellar mass (M$_*$):} It is derived from the multi-SSP
decomposition of the stellar population, spaxel-by-spaxels, performed
by the pipeline \citep{pipe3d}. This procedure provides the mass-to-light ratio
(M/L) at each spaxel, estimated as the average of the M/L of each SSP
weighted by the relative contribution in light of each one to the total
observed spectrum. Multiplying this M/L by the luminosity in each
spaxel at the considered band (in our case the $V$-band), considering
the observed flux intensity, correcting by the cosmological distance,
and the stellar dust attenuation we obtain the stellar mass within
each spaxel across the datacube. Then, the integrated stellar mass is
derived by co-adding these individual values for each spaxel, while the stellar mass
surface density (\Sst) is derived by dividing the stellar mass
within each spaxel by the area in physical quantities (pc$^2$, in our
case). For the current implementation of the analysis we adopted a \citet{salpeter55}
Initial Mass Function (IMF)
with the standard limits in stars masses (0.1-100 M$_\odot$). The errors
in the derivation of the stellar masses are dominated by a combination
of photometric calibration errors in the optical data \citep[$\sim$0.05 dex][]{sanchez16}, 
and the intrinsic uncertainties in the derivation of this quantity by
the adopted method \citep[$\sim$0.1 dex][]{pipe3d}, rather than the statistical photon noise
(that is more or less homogenized due to adopted spatial binning procedure).
When adding this later error the typical uncertainty for \Sst is of the order of $\sim$0.15 dex.

{ Star-formation rate (SFR):} The SFR is derived spaxel-by-spaxel
using the scaling relation proposed by \citet{kennicutt89} between
this quantity and the H$\alpha$ luminosity, SFR[M$_\odot$/yr]$=8\times10^{-42}$
L$_{\mathrm H\alpha}$[erg/sec] (again, adopting a Salpeter IMF), for those spaxels compatible with being ionized by { young OB stars, i.e., tracing}
star-formation (SF). { To determine which spaxels are dominated by ionization by young stars }, we use the information provided by
moment analysis of the emission lines, in particular the emission
line intensities of the \Ha, \Hb, \oiii\ \lam5007,\lam4959 and \nii\
\lam6548,\lam6584. We adopted the ionizing classification scheme
proposed in \citep{ARAA}, in which a spaxel is consider to be ionized
by SF if it is located below the \citep{kewley01} demarcation line in
the classical BPT diagnostic diagram \citep[i.e., the one involving
the \oiii/\Hb\ vs \nii/\Ha\ line ratio][]{baldwin81} and it has an
EW(H$\alpha$)$>$6\AA. Then the H$\alpha$ luminosity is derived based
on the observed flux intensity, correcting by the cosmological
distance, and the interstellar medium dust attenuation \citep[derived
from the \Ha/\Hb\ line ratio, assuming a][ extinction law]{cardelli89}.
Finally, using the scaling relation outlined before and co-adding
across the FoV of the instrument, we derive the integrated SFR.
{ Following \citet{sanchez20}, no signal-to-noise cut was applied to the emission lines
  along this process, since this could bias the results, in particular for the
  detection of low-intensity and low-EW diffuse ionized regions. We rather prefer to propagate
the errors. However, for the SF-regions, the combined requirement of
a large EW(H$\alpha$) and a physically reliable dust attenuation,
impose an implicit S/N in H$\alpha$ well above 3$\sigma$}.
Dividing the derived SFR by the area in each spaxel yields the SFR surface density (\Ssfr).
Like in the case of the stellar masses, the individual errors were derived propagating 
both the statistical errors (errors in the derivation of H$\alpha$ and the dust attenuation) and the photometric calibration ones. The typical error is of the order of $\sim$0.10 dex 
for the full sample. 

{ Molecular gas mass (M$_{\rm mol}$):} We do not have a spatial or
integrated estimation of the molecular gas content for the entire
CALIFA sample { (only $\sim$51\% galaxies have molecular gas masses
  estimated based on CO observations, as indicated below)}. However,
we can use the ISM dust attenuation, derived spaxel-by-spaxel as
described before, as a proxy based on the dust-to-gas calibrator
recently proposed by \citet{jkbb20}, defined as
$\Sigma_{\rm gas} [{\rm M_\odot pc^{-2}}]= 23 A_{\rm V}[{\rm mag}]$
(including both atomic and molecular gas).  Following
\citet{sanchez20}, we use a newly proposed functional form for this
calibrator,
$\Sigma_{\rm mol} [{\rm M_\odot pc^{-2}}]= 10^{1.37}\ (A_{\rm V}[{\rm
  mag}])^{2.3} $, presented in \citet{jkbb21a}, that reproduces the
observed molecular gas masses in a better way { (i.e., with a
  more linear one-to-one correlation between the CO based and the estimated
  molecular masses, and a lower number of outliers )}.  Considering the physical area in each spaxel, and
integrating over the FoV of the instrument we derive the molecular gas
of each galaxy. The errors in this estimation of the molecular mass
are fully dominated by the inaccuracy of the adopted calibrator.
Based on the dispersion around the one-to-one relation between the
M$_{\rm mol}$ estimated from CO measurements and based on this
calibrator \citep[$\sim$0.3 dex,][]{jkbb20}, and considering the
typical error in the CO-derived values (Sec. \ref{sec:co}), we assume
an average error of $\sim$0.25 dex for all the sample { (this is, the result
  of quadratically subtract the dispersion around the one-to-one relation with the typical error of the CO-derived values)}.

{ The analysis of the optical data described here was performed for
  each individual CALIFA galaxy twice. First, using the original
  CALIFA datacubes with the intrinsic spatial resolution of PSF
  FWHM$\sim$2.5$\arcsec$.  Then, it was repeated using a spatially
  degraded version of the datacubes, that match the resolution of the
  CO-maps, as described in more detail the next section.}


\subsection{CO datasets}
\label{sec:co}

The molecular gas estimates for the entire CALIFA sample are based on a
calibrator that may present possible biases (as calibrated using a
limited sample of observations) and secondary dependencies \citep[like
a trend with metallicity, e.g.,][]{brinchmann:2000aa}. { The net
  effect of these biases and dependencies is that this calibrator provides
  less precise estimations of the molecular gas mass, despite the fact that
  the reported values are statistically accurate. This is directly translated
  to a larger average error, as indicated before. For this reason,} we need an
independent and more { accurate} (a priori) estimation of the molecular
gas content in our sample of galaxies {,  to cross-validate the results derived from this calibrator}. With this in mind, we compiled
one of the largest dataset of CO derived molecular gas available on an
IFS galaxy survey, under the framework of the EDGE collaboration (PI:
A.Bolatto \footnote{\url{http://www.astro.umd.edu/EDGE/}}). This
collection is based on two datasets that we label hereafter as EDGE
and APEX data:

{ (i) EDGE data:} The EDGE database constitutes the first
exploration of the spatially resolved distribution of the $^{12}$CO(1-0)
at 250 MHz
(and $^{13}$CO(1-0) at 500 MHz) transition observed with the Combined Array for
Research in Millimeter-wave Astronomy (CARMA) antennae
\citep{bolatto17}. Originally, the EDGE-CALIFA collaboration mapped
177 infrared-bright CALIFA galaxies with the CARMA E-configuration
($\sim$9$\arcsec$ resolution). Of these, 125 galaxies with the highest S/N were
observed with the D-configuration, with both D and E configurations sampling the $^{12}$CO(1-0)
transition. A final resolution of 4.5$\arcsec$ is achieved when
combining the D+E configurations, which corresponds to a physical
resolution of $\sim$1.3 kpc at the average redshift of the survey
(i.e., very similar to the one achieved by ALMaQUEST). From these
observations, the molecular gas mass density
in each resolution element was derived over a FoV similar to the optical
IFS data, down to a 3$\sigma$ detection limit of $\sim$2.5 M$_\odot$/pc$^2$.
Details on the
observations, reduction, and transformation from CO to molecular gas
mass were extensively described in \citet{bolatto17}, { and summarized at the end of this sub-section}. This dataset was
extensively exploited in different articles
\citep[e.g.][]{utomo17,colombo18,levy18,leung18,dey19,jkbb20}.

The EDGE collaboration has recently created a database
(\texttt{edge\_pydb}), with the main goal to provide a homogeneous
dataset of maps combining both the dataproducts derived form the
optical and millimeter observations, including a flexible python
environment to allow the exploration of the data. Throughout this article
we make use of this database, which will be presented in detail
somewhere else (Wong et al. in prep.).  We summarize here how this
database was created, focused on the aspects more relevant to the
current exploration. First, both the CALIFA datacubes (with a PSF FWHM$\sim$2.5$\arcsec$ and the CARMA data (with an average beam size of $\sim$4.5$\arcsec$)
were degraded to the worst spatial resolution of the CO data ($\sim$7$\arcsec$). Then, the analysis performed by
{\sc Pipe3D}, described in { Sec. \ref{sec:opt}}, was repeated over the degraded
datacubes, obtaining similar physical quantities (but at a lower
resolution). Then, the maps of these different quantities were
re-sampled in a regular square grid of 3$\arcsec$, after
registering them to the CARMA WCS. This procedure limits
the possible co-variance between adjacent spaxels, which for
a sampling of 1$\arcsec$ (the original sampling of the CALIFA data),
and the adopted final beam FWHM of 7$\arcsec$,
could be considerably large. 


Finally, the resulting { spatially re-sampled }maps are
stored in a set of tables easily accessible through a python
interface. 
We will refer to these individual measurements as
line-of-sights (LoS), since they do not correspond either to the original
CALIFA spaxels or to the original CARMA CO datacubes. { We should recall that the original sampling of both datasets was 1$\arcsec$/spaxel, while the new pixel size is 3$\arcsec$, more consistent with the requirements of the sampling theorem.}. A total of
15,512 independent LoS are included in the final database
with simultaneous measurements of \Sst, \Sgas and \Ssfr, the largest dataset of this type to date. The three parameters were corrected for inclination, adopting the measurements based on our own derivations of the ellipticity and position angle for the eCALIFA galaxies \citep{carlos19,lacerda20}. In the case of
\Ssfr the database adopted an IMF different than Salpeter. For consistency we recalculate the SFR using the same procedure described for the CALIFA dataset, using the required inputs stored in the \texttt{edge\_pydb} database.

{ (ii) APEX data:} It comprises an homogenized compilation of 512
galaxies with molecular gas measured in an aperture of diameter 26.3",
described in \citet{colombo20}.  This compilation comprises new
observations of CALIFA galaxies performed using the Atacama Pathfinder
Experiment 12\,m sub-millimeter telescope (APEX;
\citealt{guesten2006}), covering the $^{12}$CO(2-1) transition at
230\,GHz.
Contrary to the CARMA observations, the only
additional selection criterion was that the objects were accessible by
the telescope (DEC$<$30$^{\circ}$), avoiding an overlap with the subsample already observed with CARMA. A total of 296 galaxy centers and
39 off center observations were obtained with good quality, of which only the first
ones were used in this article. Details on the observing strategy,
data reduction and calibration, and the procedure to transform the CO
intensities to molecular gas mass are given elsewhere \citep{colombo20}. In addition to these new observations a re-analysis of
the 177 CARMA dataset was performed to match the aperture and beam shape to that of the APEX data, providing an additional set of
aperture-matched estimates of the molecular gas. The final sample
of 512 galaxies, 430 have an estimation of the molecular gas mass within
the considered aperture ($\sim$1/3 of the area covered by the CALIFA
optical observations {, and only a fraction of the disc}). Of them, 333 correspond to well detected sources above a 3$\sigma$ detection limit of $\sim$2
M$_\odot$/pc$^2$. The remaining ones are considered as upper limits.

For the EDGE dataset, that is based on the CO(1-0) transition, it was used
the constant CO-to-H2 conversion factor $\alpha_{1-0}$=2 10$^{\rm 20}$ cm$^{-2}$
(K km s$^{-1}$)$^{-1}$ proposed by \citet{bolatto13},
as described in \citep{bolatto17}. However, for the CARMA data, it was
required an additional correction factor { of 0.7 \citep{leroy13,saintonge17}} since it was mapped the CO(2-1)
transition, as described in \citet{colombo20}.
The statistical errors associated with the individual uncertainties of each observation were
propagated to estimate the errors in the molecular gas masses and \Sgas in both datasets. In the case of the EDGE dataset the errors were already included in the \texttt{edge\_pydb} database, showing a typical error of $\sim$0.17 dex for the $\sim$12,000 LoS detected above a 2$\sigma$ detection limit. In the case of the APEX dataset the statistical errors are much lower than the calibration ones \citep[Sec. 2.2 of][]{colombo20}. Once considered them, the typical error is $\sim$0.15 dex for the $\sim$400 galaxies detected
above 2$\sigma$. In addition to the CO derived molecular gas, we extracted the aperture-matched stellar mass and SFR derived from the optical observations, as described in \citet{colombo20} and Colombo et al. (in prep.), { to build-up the final dataset for the APEX subsample.}

\subsection{Main properties of the sample}
\label{sec:main}

\begin{figure}
 \minipage{0.99\textwidth}
 \includegraphics[width=8.5cm]{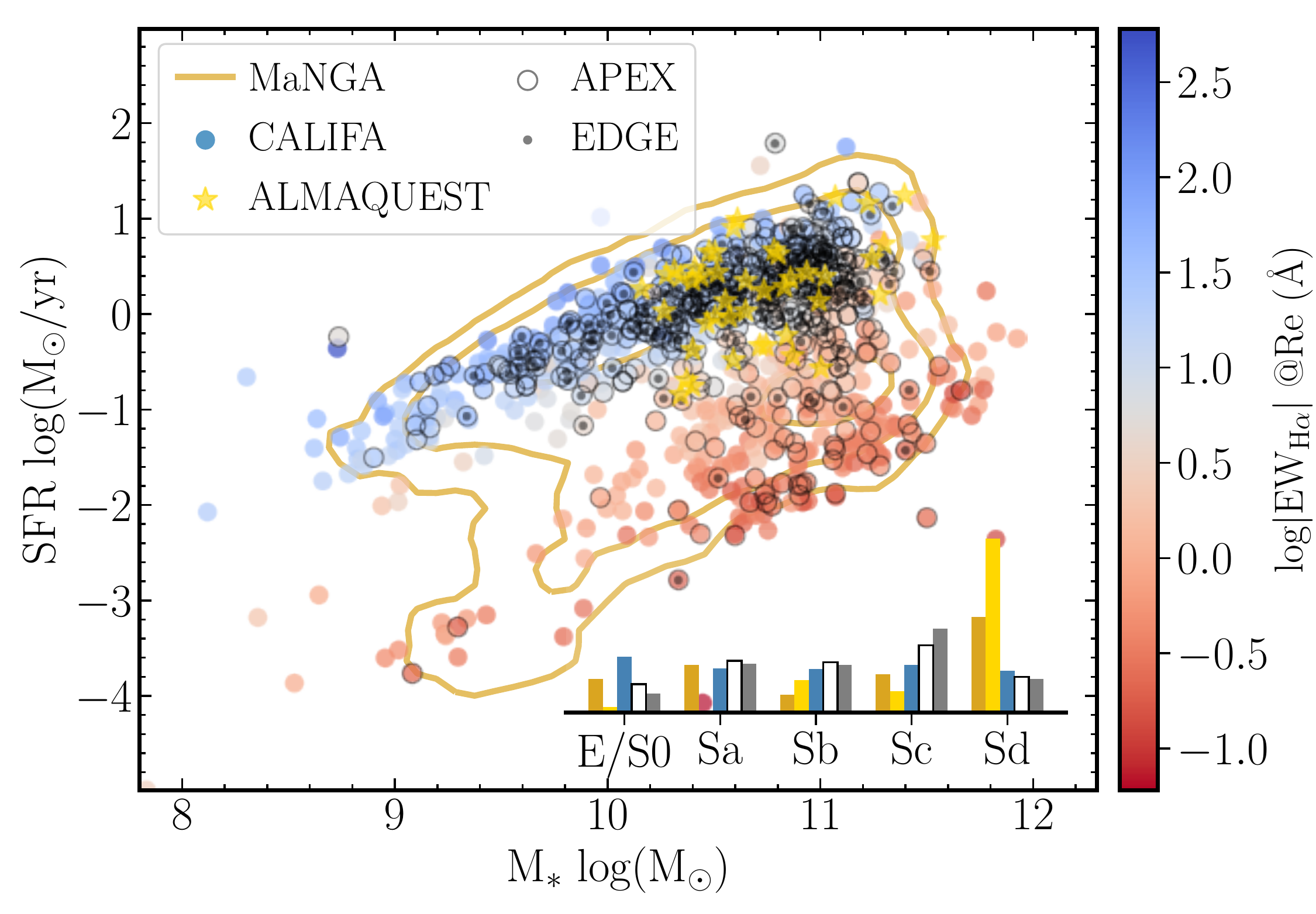}
 \endminipage
 \caption{Distribution of the explored galaxies along the SFR-M$_*$
   diagram. Solid-circles correspond to the extended CALIFA sample,
   color-coded by the EW(H$\alpha$) at the effective radius of each
   galaxy. Those galaxies with integrated CO-observations (APEX
   sub-sample) are represented with an additional black open-circles,
   while those with spatially resolved CO-observations (EDGE
   sub-sample) are represented with an additional smaller grey
   solid-circles. For comparison purposes we include the distribution  
   for the { full sample of } galaxies explored by ALMaQUEST
   \citep[yellow solid stars][]{lin20}, extracted from
   the MaNGA survey \citep{manga}. The distribution of the full MaNGA
   sample, DR15 distribution, is represented by successive
   golden-color density contours (encircling 95\%, 40\% and 10\% of
   the sample). The inset shows the morphological distribution of
   galaxies for the different samples (in terms relative), using a
   similar color-coding: (i) CALIFA in blue, with (ii) APEX in white, including
   an additional solid black-line, and (ii) EDGE a shaded grey
   pattern; (iii) ALMaQUEST in yellow and (iv) MaNGA-DR15 in golden
   colors.}
 \label{fig:sample}
\end{figure}

Figure \ref{fig:sample} shows the distribution of the full extended CALIFA
sample, together with the EDGE and APEX subsamples, along the
SFR-M$_*$ diagram, color-coded by the characteristic EW(H$\alpha$)
(i.e., that at an annulus located at the effective radius, R$_e$, of the galaxy). { We should re-call here that we will analyse each sub-sample
  independently, since they cover the space of parameters at different
  spatial resolutions and/or apertures, to narrow-down their effects in
  the explored relations. Therefore, it is needed to understand how each
of them covers the space of parameters of galaxies in the nearby universe.}
For comparison purposes we have
added to the figure the distribution of the MaNGA DR15 \citep{sdss_dr15}
galaxies (the largest IFS galaxy survey in terms of number of objects),
and the distribution of galaxies explored by the ALMaQUEST survey (Lin
in prep.). {  This later was included since a subset of this sample was
used by \citet{lin19} and \citet{ellison21a} to perform a similar exploration as the one
presented here.}

It is evindent in the figure that the extended CALIFA sample covers a
similar region of the diagram as the MaNGA DR15 one, despite the fact
that it comprises 1/4th the number of galaxies. The stellar mass
regimes covered by the two samples are very similar, with most of the
galaxies located between 10$^{8.5}$ and 10$^{11.5}$ M$_\odot$ in both
cases \citep[as already noted by][]{sanchez17a,ARAA}.  Furthermore, most galaxy types in terms of
the star-formation properties are well represented by the current
adopted sample. Following \citet{ARAA}, if we define star-forming galaxies (SFGs) as those
with a characteristic EW(H$\alpha$)$>$6\AA { (0.78 dex in logarithm, i.e., the transition between red and blue colors in Fig. \ref{fig:sample}) }, and without a trace of AGN
\citep[i.e., excluding the 34 AGN candidates described
in][]{lacerda20}, we retain 532 galaxies. { Thus, the selection procedure
  is the same adopted in Sec. \ref{sec:opt} to select star-forming regions/spaxels.}
The remaining $\sim$350
galaxies would be either green-valley or retired galaxies \citep[e.g.,
][]{sta08,cid-fernandes10}. This ratio of SFGs to RGs is very
similar to that expected for a representative population of galaxies in
the nearby-universe \citep[e.g.][]{na2010,sanchez18b}.

If we consider the subsample of galaxies with CO observations, we do find
two different behaviours: (i) the APEX subsample presents a good
coverage of the average population in terms of stellar masses
(covering the same range as the full extended CALIFA sample) and the
star-formation stage (with 251 SFGs out of 512 objects); and (ii) the
EDGE subsample is restricted to the narrower stellar mass regime
(M$_*>$10$^{9.4}$M$_\odot$), and more biased towards the SFGs
($\sim$110 of 126 objects). This would not be an important bias if
the properties of star-forming regions were similar in both
star-forming and retired or green valley galaxies.


The morphological distributions of all the samples and subsamples
included in the diagram are presented in the inset in Figure
\ref{fig:sample}. It is evident that the CALIFA sample covers all
the different morphological types, following a more homogeneous
distribution than any of the other samples, including the MaNGA
DR15. The bias in the morphological distribution of the MaNGA sample
with respect to the expected distribution for galaxies in the nearby
universe has been discussed before \citep[e.g.][]{sanchez18},
and it is beyond the scope of the current article. What it is more
relevant is that both the APEX and EDGE sub-samples cover a wide range
of galaxy morphologies, with a general trend towards including more late-
than early-type galaxies than the original CALIFA sample from which
they were drawn. This is understandable since the EDGE sample is
clearly biased towards SFGs due to their selection criteria (bright
infrared galaxies), and therefore towards later types. However,
both samples still include some early-type galaxies and a considerable
number of early-spirals (Sb/Sa).

When comparing with the observed distribution for the ALMaQUEST
compilation, the first difference to highlight is the { lower}
number of galaxies: 46 objects in the full sample \citep{lin20}, with
only 14 and 28 of them used in their explorations of the resolved
relations \citep{lin19,ellison21a}.  Furthermore, it is clear that
their sample { comprises a larger faction of SFGs} than either the
APEX or EDGE subsamples, with a { deficit} of retired galaxies {
  \citep[although it covers the Green Valley and comprises several
  retired regions e.g.][]{ellison21a,ellison21b}}.  { It also
  covers a narrower range of stellar masses ($>$10$^{10}$M$_\odot$)
  and morphologies. The ALMaQUEST sample has a larger fraction
  late-type galaxies (Sd in most of the cases), with a deficit of
  early-spirals (in particular, no Sa) and no ellipticals or S0
  galaxies.} This highlights the importance of revisiting the
exploration of the described global and local relations with a
different sample to determine the importance (or not) of these
{ selection effects}.

It is relevant to highlight that our sample contains a similar number of galaxies
as the largest prior explorations of the molecular gas content of galaxies
in a redshift range ($z<$0.1)
comprise similar number of objects as the one studied here. For
instance, the COLD GASS survey, the reference survey in this field
\citep[][]{saintonge2011}, mapped 532 SDSS galaxies using the IRAM-30m
telescope. This number of objects is similar to the one comprised by
our APEX subsample. It is worth noting that the beam size of their
observations is very similar too, so, for galaxies at a similar
redshift their molecular gas content is restricted to the same galaxy
regions.

{ In summary,} the adopted sample and subsamples are
well placed to study the general behaviors of the explored relations
due their main properties: (i) they cover a relatively narrow redshift
range and a low average redshift ($z\sim$ 0.015), which guarantees a
relatively small and quite homogeneous physical resolution ($\sim$800
pc); (ii) the FoV of the IFS explorations ($>~$1 arcmin$^2$), that,
together with the diameter selection guarantees a large covered
optical extent of the galaxies (up to $\sim$2.5 Re) with a good
spatial sampling; (iii) the wide range of stellar masses,
star-formation stages and galaxy morphologies sampled; and finally, (iv)
the spatial coverage and physical resolution of the two datasets
of CO observations, that are similar to those of most recent explorations.

\begin{figure*}
 \minipage{0.99\textwidth}
 \includegraphics[width=\linewidth]{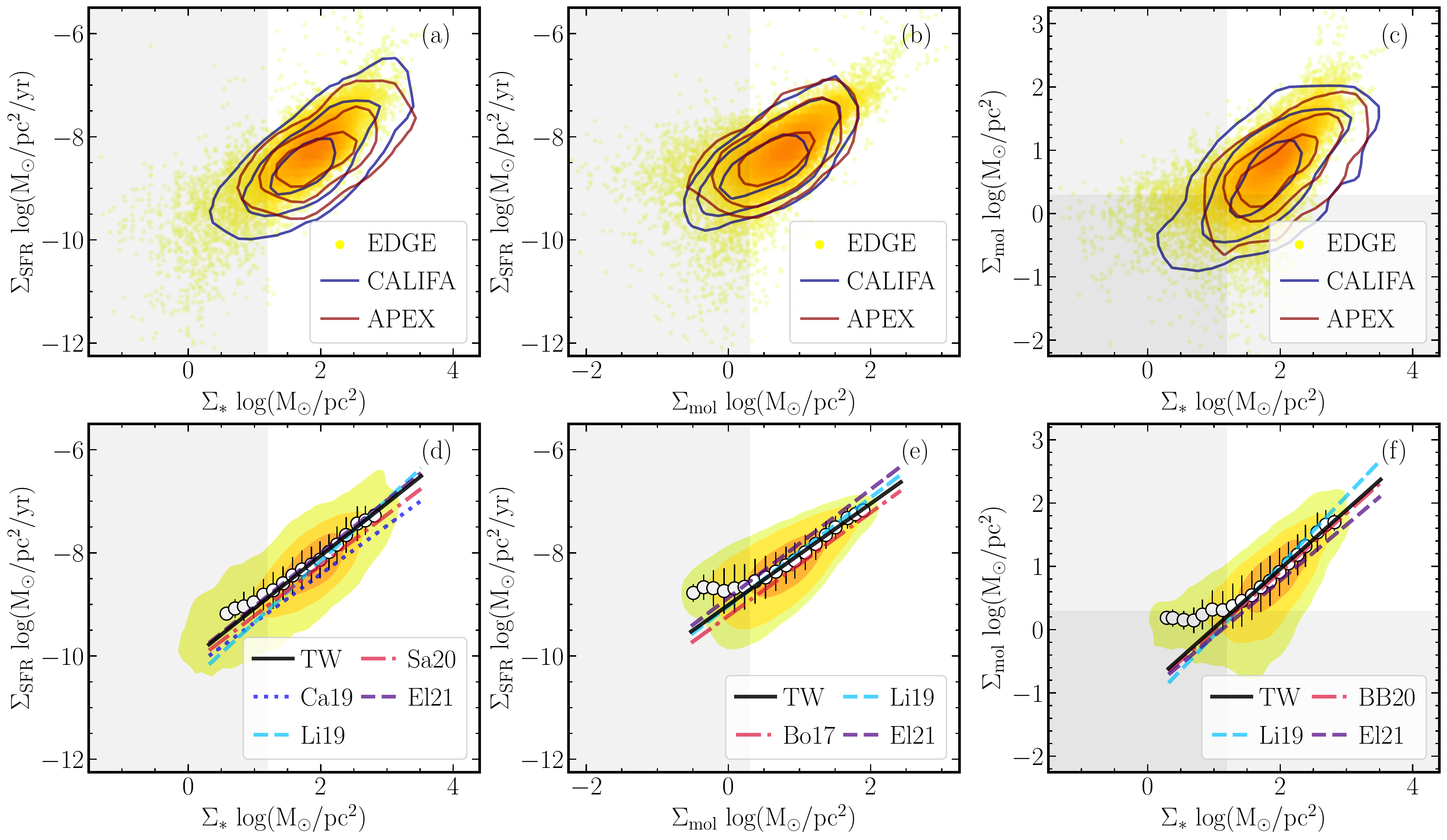}
 \endminipage
 \caption{(a) Distribution of \Ssfr along \Sst for the
   $\sim$15,000 LoS across the $\sim$100 galaxies observed
   by the EDGE survey (solid circles, color-coded by the density of
   points). Blue and red contours show the average \Ssfr vs the
   average \Sst for the full CALIFA sample ($\sim$900 galaxies), and the APEX sampled
   objects ($\sim$400 galaxies).  In the case of CALIFA, the values correspond to the
   average quantities across an effective area (as described in Sec. \ref{sec:ana}), while in the case of APEX they correspond to the average
   values within the beam of the antenna. Each consecutive contour
   encircles  95\%, 40\% and 10\% of each considered sample. Only those LoS (or individual measurements) compatible
   with ionization dominated by SF, as described in the text, have been included in this plot; (b) and
   (c) panels show similar distributions for the \Ssfr vs. \Sgas and \Sgas vs. \Sst diagrams.
   Lower panels show the same distributions for the same parameters shown
   in the corresponding upper panel for the EDGE data, represented as filled contours (with the
   same encircled number of points as the contours in the upper panels). White solid circles
   in each lower panel show the mean values in bins of 0.15 dex in the corresponding
   parameter represented in the X-axis, with error bars indicating the standard deviation
   around each mean value. The solid black line shows the best fitted linear regression to
   these solid circles (values within the shadow region, indicating the regime affected
   by detection issues, were excluded from the fitting). These regressions, labelled as TW (for This Work), correspond
   to the rSFMS ($d$ panel), rSK ($e$ panel) and rMGMS ($f$ panel), relations. For comparison
   purposes, we have included some of the most recent derivations of these relations
   extracted from the literature, including: \citet{bolatto17} (Bo17), \citet{lin19} (Li19),
   \citet{mariana19} (Ca19), \citet{jkbb20} (BB20), \citet{sanchez20} (SS20) and \citet{ellison21a} (El21). }
 \label{fig:2D}
\end{figure*}

\section{Analysis and Results}
\label{sec:ana}

One of the main goals of the current study is to determine how the
global intensive relations described for SFGs by the galaxy-wide
average \Ssfr, \Sst and \Sgas are connected with the local/resolved
relations found between the same quantities at kiloparsec scales.  To
do so, we follow \citet{sanchez20}, and derive the
described global intensive quantities by dividing each extensive
quantity (SFR, M$_*$ and M$_{\rm mol}$ by (i) the effective area of each
galaxy (defined as the area within 2r$_e$, i.e.,
A$_e$=4 $\pi$ r$_e^2$) for the case of the CALIFA sample, and (ii)
the beam area of each CO observation in the case of the APEX
subsample \citep[that corresponds in average to $\sim$1 R$_e$ of the galaxies][]{colombo20}. For the latter, the choice of area is purely justified by
the nature of the observations. In the case of the CALIFA sample, we
choose an area representative of the coverage of the IFS data and
attached to the properties of each galaxy. However, as already
discussed in \citet{mariana19} the use of any area of the order of the
size of the FoV of the instrument would provide similar results.
{ The EDGE sub-sample comprises already extensive quantities, as
  it provides with the spatial resolved \Ssfr, \Sst and \Sgas for the different
  LoS included in the database (Sec. \ref{sec:co}. Therefore, no normalization by
  the area is required). Finally, } 
we apply an inclination correction using the same parameters adopted in the
correction of the EDGE dataset (Sec. \ref{sec:co}). This
procedure provides a single set of \Ssfr, \Sst and \Sgas for the
subset of SFGs extracted from the CALIFA (532 galaxies) and APEX
datasets (251 galaxies). The distribution of these quantities can
be then compared directly with kiloparsec scale values provided by
the EDGE dataset.

\begin{table*}
\caption[]{Results of the analsys of the rSFMS, rSK and rMGMS scaling relations}
\label{tab:2D}
\begin{tabular}{llrrrrrrr}
\hline
  \multicolumn{1}{c}{Relation} & \multicolumn{1}{c}{Reference} & \multicolumn{1}{c}{$\beta$} & \multicolumn{1}{c}{$\alpha$} &  \multicolumn{1}{c}{r$_c$} & \multicolumn{1}{c}{$\sigma_{\rm obs}$} & \multicolumn{1}{c}{$\sigma_{exp}$} &
  \multicolumn{1}{c}{\# galaxies} &
  \multicolumn{1}{c}{\# SFA/SFG}\\
  \hline
  rSFMS & EDGE   & -10.10$\pm$0.22 & 1.02$\pm$0.16 & 0.68 & 0.266 & 0.190 & 126 & 12667\\
        & CALIFA & -10.27$\pm$0.22 & 1.01$\pm$0.15 & 0.85 & 0.244 & 0.192 & 941 & 533\\
        & APEX   &  -9.78$\pm$0.30 & 0.74$\pm$0.21 & 0.76 & 0.226 & 0.211 & 512 & 251\\
      & \citet{lin19}        & -11.68$\pm$0.11 & 1.19$\pm$0.01 &    & 0.25 &   & 14 & 5383$^{*}$
  \\
      & \citet{mariana19}    & -10.48$\pm$0.69 & 0.94$\pm$0.08 & 0.62 & 0.27 &  & 2737 & $\sim$500K\\
      & \citet{sanchez20}    &-10.35$\pm$0.03 & 0.98$\pm$0.02 & 0.96 & 0.17 &   &1512 & $\sim$3M\\
      & \citet{ellison21a}   &-10.07$\pm$1.44 & 1.03$\pm$0.17 & 0.57 & 0.28-0.39 &   & 28 & $\sim$15035$^{*}$\\      
\hline
rSK  & EDGE   & -9.01$\pm$0.14 & 0.98$\pm$0.14 & 0.73 & 0.249 &  0.216 & 126 & 12667\\
     & CALIFA & -9.01$\pm$0.16 & 0.95$\pm$0.21 & 0.77 & 0.293 &  0.297 & 941 & 533\\
     & APEX   & -8.84$\pm$0.24 & 0.76$\pm$0.27 & 0.70 & 0.294 &  0.228 & 512 & 351\\
     & \citet{bolatto17}    & -9.22            & 1.00          &      &      & & 104 & $\sim$5000\\
     & \citet{lin19}        & -9.33$\pm$0.06   & 1.05$\pm$0.01 &      & 0.19 & & 14 & 5383$^{*}$\\
     & \citet{ellison21a}   & -8.87$\pm$0.66 & 1.05$\pm$0.19 & 0.74  & 0.22-0.32 &   & 28 & $\sim$15035$^{*}$\\  
\hline      
rMGMS & EDGE  & -0.91$\pm$0.16 & 0.93$\pm$0.11 & 0.68 & 0.218 & 0.209 & 126 & 12667\\
      & CALIFA& -1.12$\pm$0.27 & 0.93$\pm$0.18 & 0.74 & 0.276 & 0.288 & 941 & 533\\
      & APEX  & -0.70$\pm$0.37 & 0.73$\pm$0.24 & 0.73 & 0.234 & 0.212 & 512 & 251\\
      & \citet{lin19}        & -1.19$\pm$0.08  & 1.10$\pm$0.01 &    & 0.20 & & 14 & 5383$^{*}$\\
      & \citet{jkbb20}       & -0.95           & 0.93          &    & 0.20 & & 93 & $\sim$5000\\
      & \citet{ellison21a}   & -0.99$\pm$0.13  & 0.88$\pm$0.15 & 0.72 & 0.21-0.28 &   & 28 & $\sim$15035$^{*}$\\
\hline
\end{tabular}
\begin{minipage}{\textwidth}
  Best fitted intercepts ($\beta$) and slopes ($\alpha$) for the
  different resolved and global (intensive) relations explored in this
  article (rSFMS, rSK and rMGMS) and the different explored dataset
  (EDGE, CALIFA and APEX). Some of the most recent derivations
  extracted from the literature have been included for comparisons. In
  addition we include the correlation coefficients ($r_c$), the
  standard deviation around the best fitted relation
  ($\sigma_{\rm obs}$) and the expected standard deviation due to the
  error propagation ($\sigma_{\rm exp}$), together with the number of
  galaxies in each sample and the number of star-forming areas (SFA)
  or star-forming galaxies (SFG) used in the derivation of the
  considered relation. $(*)$ We should note that in the case of
  ALMaQUEST we list the number of spatial elements reported in the quoted articles.
  However, those are not fully independent measurements, since they adopted the
  original MaNGA sampling of 0.5$\arcsec$/size spaxels. For a PSF FWHM of $\sim$2.5$\arcsec$,
  the true number of fully independent measurements (i.e., LoS) would be $\sim$25 lower than
  the reported values.
\end{minipage}
\end{table*}

Figure \ref{fig:2D}, top panels, shows the comparison between
distributions along the \mbox{\Ssfr-\Sst (left)}, \mbox{\Ssfr-\Sgas}
(middle) and \mbox{\Sgas-\Sst} (right) diagrams of the global
intensive quantities derived for the CALIFA and APEX datasets (as
described before) together with the corresponding spatially resolved
quantities derived for the EDGE dataset as directly extracted from the
database. It is clear that the distributions of the resolved and
global parameters cover the same range of values, following the same
trends. A simple $\chi^2$-test comparing each density distribution
indicates that they are statistically indistinguishable when comparing
the regions encircled by a 95\% of the points (first contours in the
upper figures). Significant differences appear only when the
comparisons are restricted to the peak of the distributions (region
encircling $\sim$20\% of the points), in particular when comparing the
APEX dataset with the other two. This is expected since the APEX CO
observations are restricted to the central $\sim$26$\arcsec$ of the
galaxies { ($\sim$8 kpc at the average redshift of the sample)},
and therefore their distribution is slightly shifted towards higher
values in the three distributions.  However, the effect is
insignificant when comparing the full distributions. This indicates
that the range of average values covered by the CALIFA and APEX
distributions is very similar to the range of spatially resolved
values covered with each galaxy. Finally, the distribution for the
EDGE dataset enhibits the strongest differences with respect to the
other two only in the regimes outside the contour encircling the 95\%
of the points. Above that density limit this dataset follows in
general the same trends as the other two. In summary, this analysis
shows that the distributions of the global intensive parameters are
the same as those of the resolved ones, as suggested by previous more
limited explorations \citep[e.g.,][]{mariana19,ARAA,sanchez20}. {
  Furthermore, it seems that the caveats regarding the possible biased
  in the EDGE sub-sample towards SFGs are not relevant for the
  exploration of these distributions.}

Now that we have determined the similarity between the intensive global and
resolved distributions, we now characterize the relations among them
(i.e., the so-called rSFMS, rSK and rMGMS relations). To do so, we
follow a similar procedure as the one performed by previous studies
in the characterization of global relations \citep[e.g.][]{sanchez17a,jkbb20}.
First, for each distribution, we derive the median values within
a set of consecutive bins of 0.15 dex in the parameter shown in the x-axis
of each panel (\Sst first and last, and \Sgas in the middle one), within
the plotted range of values. These bins
are restricted to the regime of points encircled by the 95\% density contour,
in order to exclude outliers. Finally, in the case of the EDGE dataset
we restricted our fit to those bins within the 3$\sigma$ detection
limits \citep[as reported by][]{jkbb20}. These limits are maybe
a bit conservative since they refer to the original detection limits
of the CO observations, prior to the re-sampling described in Sec. \ref{sec:sample},
that has certainly increased the signal-to-noise (S/N, per resolution element). The areas below these detection limits
are shown as shaded regions in the different panels of Fig. \ref{fig:2D}.
Despite the possible overestimation of the exact value, 
the individual points in this regime are detected with
a low SN, what it is reflected in an increase of the scatter (more
clearly appreciated in the \Ssfr-\Sgas and \Sgas-\Sst distributions),
and a deviation of the average distributions from the general trend.

Fig. \ref{fig:2D}, bottom panels, illustrates this procedure. The
shaded contours show the density distribution of EDGE points in the
different panels, and the white solid-points the binned values (with
errorbars indicating the standard deviation around the the
corresponding value). The solid line (labelled TW, for {\it This Work}), describe the best
fitted linear relation between each pair of parameters. This relation
was performed using a simple linear regression for the white points (weighted by the plotted errorbars, that corresponds to the standard deviation of the distribution),
taking into account the restrictions indicated before. In other words,
the values were fitted to the expression
\begin{equation}
  {\mathrm{log}} {\mathrm P_j} = \beta + \alpha {\mathrm{log}} {\mathrm P_i}
\label{eq:log}
\end{equation}
\noindent that
corresponds to the power-law $P_j = \beta'\ {\mathrm P_i}_*^\alpha$,
where $\beta$=log($\beta'$). In order to consider the individual
errors, we perform a Monte-Carlo iteration perturbing the original data
within their errors, and repeating the full analysis 100 times. The
reported relation corresponds to the average of the MC iteration.

Table \ref{tab:2D} shows the results of this fitting procedure for the
different datasets explored in this study (CALIFA, APEX and EDGE),
together with the most recent derivations extracted from the
literature. For each of the analyzed relations we present the
intercept ($\beta$) and slope ($\alpha$) of the relations, together
with the correlation coefficient (r$_c$) of the original pair of
correlated values, and the standard deviation of the y-axis parameter
once subtracted the best fitted power-law ($\sigma_{\rm obs}$). For comparison
purposes we include the standard deviation { that would be expected
  if the parameters fulfilled perfectly well the derived relations
  and all the source of dispersion is due to the individual errors} ($\sigma_{\rm exp}$). It is important
to indicate that literature data have been transformed to our current
units (in particular area measured in pc$^2$). As already discussed in
\citet{sanchez20}, the intercept in the reported relations
is not a dimensional quantity, and to shift one relation derived for
surface densities in kpc$^2$ to one in pc$^2$, the actual value of the
slope ($\alpha$) plays a role. { Finally, in the case of \citet{ellison21a},
that reported two set of values for each relation (based on two different fitting
procedures), we list the average of them (corrected to the adopted units too), with
the errors corresponding to a 25\% of the difference between them.}


Focusing on the local/resolved relations, we find good agreement with
the results presented by previous explorations. This is particularly
true for the slopes of the relations, that in all cases differ less
than one or two sigma from those extracted from the literature.  The
agreement is indeed extremely good with those recently reported by
\citet{jkbb20} for the rMGMS relation. That is not a surprise since
both quantities are derived from the same dataset, although using a
different extraction of the LoS. They adopted a less restrictive
scheme in terms of the spatial sampling than the one presented here,
too.  { The agreement is also very good with the values reported by
\citet{lin19} and \citet{ellison21a}, although there are some
appreciable differences.} The slopes reported by \citet{lin19} are in
general larger, although at about one sigma from the ones reported
here. Their reported errors for the slopes are much smaller than the
ones we find. The differences in the slope are translated to an
apparent larger discrepancy in the intercepts. We consider it
artificial since it is a consequence of the extrapolation of the
relations (with slightly different slopes) to a regime not covered by
the observed data \citep[as nicely discussed by][recently]{mariana19}.
This indeed can be appreciated in Fig. \ref{fig:2D}, lower panels,
where we have included the actual relations reported in the literature
and listed in Tab. \ref{tab:2D}, showing a very good agreement with
both the ones derived in the current study and the distribution of
points for the EDGE dataset.

{ On the other hand, the average slopes reported by \citet{ellison21a}
are more similar to the ones reported here. We should note that this
study updated the analysis of the ALMaQUEST data by \citet{lin19},
increasing the number of galaxies by a factor two (and the number of
sampled regions by a factor three). In particular, they have a better
coverage of the space of parameters, with more green valley
galaxies. As discussed in that article the actually fitting procedure
matters, with ordinary least squares (OLS) applied to the full sample
of regions providing shallower slopes than the one provided by an
orthogonal distance regression (ODR) procedure. These later ones are
more similar to those reported by \citet{lin19}. Our current analysis
is neither both of them, since it involves a prior binning. We adopted
that procedure based in previous experience, and our tests indicate
that the recovered values correspond to the real ones within the
errors (Sec. \ref{sec:sec}).} Finally, we note that the largest
difference is found for the relation presented by \citep{mariana19}
for the rSFMS, which in any case is just at one sigma of our observed
distribution (i.e., just at the edge of errorbars in the figure).

Having established that our local/resolved relations are similar
to the most recent published ones, particularly those using CO
observations of similar spatial resolution \citep[e.g.][]{lin19,ellison21a}, we continue
exploring how those relations compare to the global intensive ones. In
the case of the CALIFA dataset, both the intercepts and in particular
the slopes agree with the values found for the local
relation (EDGE) or reported in the literature (e.g., ALMaQUEST){, both listed in Table \ref{tab:2D}}. The
main difference, as indicated before, is in the errors reported for both quantities, that in
general are larger for the rSK and rMGMS relations. This is 
expected, since from the CALIFA sample we use an indirect proxy for the
molecular gas that most probably adds an extra dispersion in the two
relations involving \Sgas \citep{jkbb20}. The strongest differences are found between
the parameters reported for the different relations derived using the
APEX dataset when compared with the other two (and the literature
values). This may looks contradictory since this dataset comprises
molecular masses derived from direct CO observations. However, it is
easily understandable when considering that the errors in two
estimated coefficients are larger (in some cases considerably larger)
than those reported for the derivations based on the CALIFA and EDGE
datasets. Indeed, the actual values of the parameters are
statistically compatible with those reported for the other datasets
(and the literature) within less than one sigma difference. We
consider that the combined effect of (i) APEX being the dataset with the
lowest number of individual measurements and (ii) its slightly narrower
range of covered parameters has produced this effect. Note that although the APEX dataset includes all EDGE galaxies, it comprises only one single set of values per galaxy (\Sst, \Ssfr, \Sgas, aperture matched), not the full range of values covered by the spatially resolved EDGE data. Actually, randomly selecting a similar number of galaxies from the CALIFA dataset, restricted to the same range of parameters covered by APEX,
we can reproduce the similar slopes in $\sim$40\% of the cases.

As indicated before the errors reported by \citet{lin19} in both the
intercept and slopes of all the relations are considerably lower
than the ones reported here and in most of the studies in the literature.
\citet{sanchez20} is a counter-example. However, in that particular case
only the formal errors with respect to the averaged binned points were reported.
This could be well the case for \citet{lin19}, although they reported an ODR-fitting
for the full analyzed dataset. Another possible reason of their low reported
errors could be the strong co-variance among their points, since they
used individual pixels/spaxels of 0.5$\arcsec$ when
their actual PSF FWHM is 2.5$\arcsec$ (i.e., they sampled each resolution element
with $\sim$25 non-independent pixels). { However, this cannot be the main reason,
since in the update analysis of the ALMaQUEST data (with a larger number of measurements)
by \citet{ellison21a} they reported errors more similar to the ones reported here. Thus,
we consider that most probably they have reported just the formal errors.}

\subsection{Preference among the rSFMS, rSK and rMGMS relations}

\citet{lin19} reported a clear prevalence among the explored relations, based on the Pearson correlation 
coefficient and the standard
deviations of the residuals (i.e., once subtracted the best fitted
model). Based on their analysis the rSK is the primary relation
between the explored parameters, with a correlation coefficient
significantly larger (r$_c=$0.81), and a standard deviation slightly
lower ($\sigma=$0.19 dex), than that of the other two (r$_c$=0.76 and
0.64, and $\sigma=$0.20 and 0.25 dex, for the rMGMS and rSFMS
respectively).  Their conclusion is that the combination of the rSK
and rMGMS relations leads to the rSFMS as a by-product of the other
two correlations.{  Similar results were found in the updated analysis
by \citet{ellison21a}, in the sense that the rMGMS and rSK present a tighter
correlation (lower standard deviations) than the rSFMS.} So far, we cannot confirm their results and
conclusions. Neither for the resolved nor for the intensive global
relations explored in here we find a clear prevalence among them. For
the EDGE dataset the correlation coefficients are
essentially equal among the different explored relations, with an
r$_c\sim$0.68-0.73, as seen in Tab. \ref{tab:2D}. Although the
strongest correlation is found for the rSK (r$_c=$0.73), there is neither such a big
difference as the one reported by \citet{lin19}, nor a clear ordering
(indeed, the r$_c$ found for the rSFMS is the same as the 
one found for the rMGMS). Furthermore, the minimum standard deviation of the
residual is found for the rMGMS ($\sigma_{\rm obs}=$0.218 dex), and not for the rSK. Indeed,
this later one exhibits a standard deviation of the residual just slightly
lower than that of the rSFMS ($\sigma_{\rm obs}=$0.249 and 0.266 dex, respectively). 
For a proper comparison among the different relations is needed to consider the
expected standard deviation due to the propagation of the errors in the observed quantities and the estimated slopes ($\sigma_{\rm exp}$ in Tab. \ref{tab:2D}). When doing so
we find that the rMGMS is indeed the relation with the smallest residuals, with
a dispersion essentially dominated by the errors, with the rSFMS being the one with the largest dispersion, on the other hand (with an intrinsic dispersion $\sim$0.066 dex). Note
that we have included the third decimal to reveal any possible difference,
although we find it difficult to justify the significance of any decimal beyond
the second one.

Similar results are found for the aperture-limited and global intensive scaling relations.
In the case of the APEX dataset, the rSK relation is the one with the
weakest correlation (r$_c$=0.73), having the largest standard
deviations of the residuals ($\sigma_{\rm obs}=$0.294 dex vs. $\sigma_{\rm exp}=$0.228 dex), with the rSFMS the one
that shows the strongest one (r$_c$=0.76), with the lowest standard deviation ($\sigma_{\rm obs}=$0.226 dex vs $\sigma_{\rm exp}=$0.211 dex).

\begin{figure*}
 \minipage{0.99\textwidth}
 \includegraphics[width=\linewidth]{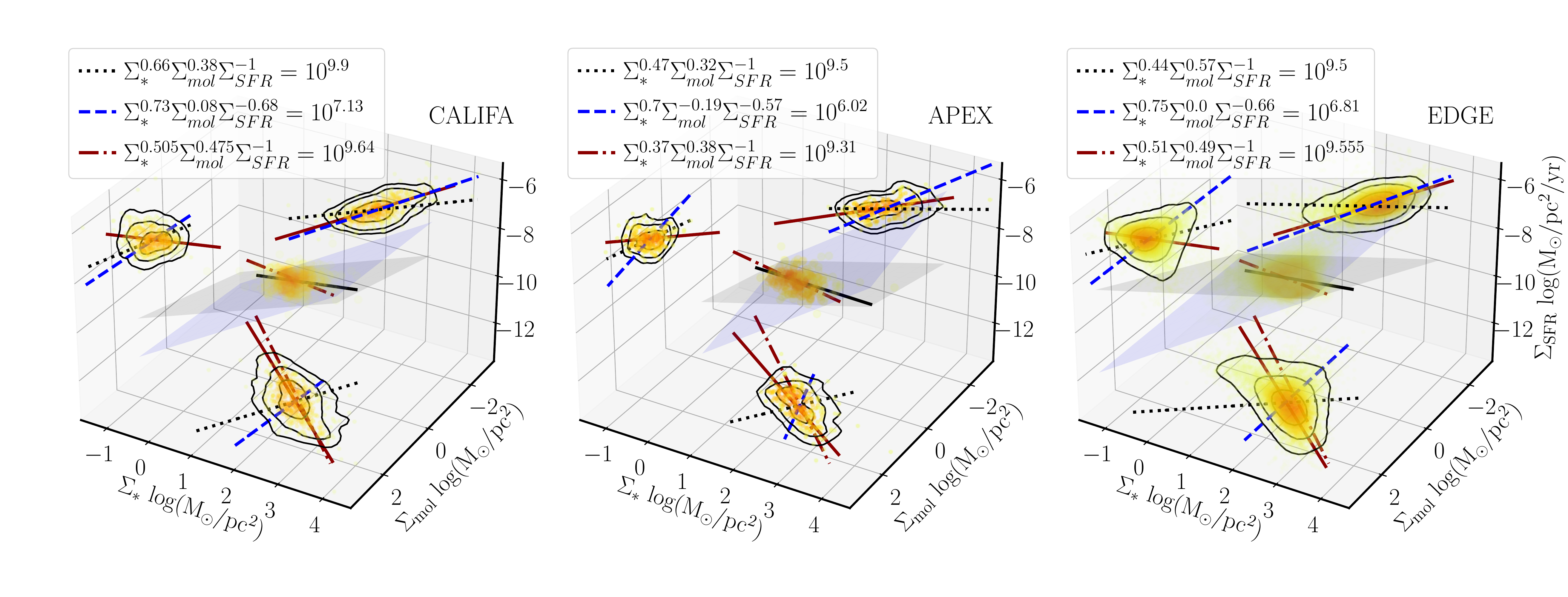}
 \endminipage
 \caption{Distribution of  \Ssfr, \Sgas and \Sst in three
   dimensional space for CALIFA (averaged across each galaxy, left
   panel), APEX (averages across the antenna beam, middle panel), and
   EDGE (individual kpc-scale LoS, right panel). Each
galaxy (CALIFA/APEX) or region within a galaxy (EDGE) is
   represented as a single solid circle, both in the 3D space and
   projected in each axis. Each circle is color-coded according
   to the local 3d density of points. Contours in each axis represent
   the regions encircling a 95\%, 40\% and 10\% of the
   points. Semi-transparent planes, and their corresponding
   projections in each axis, describe the location of the best fitted
   planes to the observed distribution of points assuming: (i) that
   \Ssfr\ depends on both \Sst and \Sgas (i.e., that those parameters
   are independent), represented by a grey plane (in 3d) and three
   black-dotted lines (in each axis), i.e., the ``Plane''-model approach, as described in the text; and (ii) that there is no {\it a
     priori} knowledge of which is the dependent or independent
   parameter, represented by a blue plane (in 3d) and three
   blue-dashed lines (in each axis). Thus, the ``PCA''-model approach, as described in the text. The estimated parameters
   describing each plane are included in the inset label. Solid
   dark-red lines in each axis correspond to the best fitted linear
   regressions to each pair of data derived using the procedure
   outlined in Fig. \ref{fig:2D}, i.e., the rSFMS, rSK and rMGMS
   relations. In addition, the average of the rSFMS and rSK relations, and their projection in the \Sgas-\Sst plane, is shown as a dashed-dotted 3D line too. This is the first functional form adopted to describe a average "Line"-model, as indicated in the text. It
   is evident from the figure that none of the planes described before
   (grey or blue), nor their projections in the axis, describes
   the distribution of points as well as the 3D dark-red solid line,
   as discussed in the text. Finally, the solid black-line, in each panel, corresponds to the best-fitted line in the three dimensional space derived using a simple linear regression. A version of this figure for different projection angles is included in the following link: \url{http://ifs.astroscu.unam.mx/CALIFA/EDGE/Scaling_Relations/}}
 \label{fig:3D}
\end{figure*}

Finally, for the CALIFA dataset the
strongest correlation is the one found for the rSFMS (r$_c$=0.85),
followed by the one for the rSK (r$_c$=0.77), that it is just slightly
larger than that of the rMGMS (r$_c$=0.74). The rSFMS is the relation with
the lowest observed standard deviation, too ($\sigma_{\rm obs}=$0.244 dex), with the rSK exhibiting the largest one ($\sigma_{\rm obs}=$0.294 dex). However, in this case, when
comparing with the expected standard deviations due to the errors we observe that indeed both the rSK  and rMGMS are completely dominated by them ($\sigma_{\rm obs}=$0.297 dex and $\sigma_{\rm exp}=$0.288 dex, respectively).

The relative weakness of the relations involving \Sgas\ for CALIFA
(compared with the rSFMS one) could be attributed to some extent to
the fact that the gas mass is derived using an indirect
proxy (with the largest errors with respect to the rest of the involved analyzed quantities). 
However, this does not change (or affect) the fact that the
correlation coefficient derived for the rSFMS in this dataset is the
strongest among all the explored relations. Indeed, the strength of
the resolved rSFMS reported by \citet{lin19} { and \citet{ellison21a} are among} the weakest
found in the literature { (see Tab. \ref{tab:2D})}. For instance, \citet{mariana16} and
\citet{mariana19}, based on the exploration of the star-forming
regions in the CALIFA and MaNGA-DR15 dataset respectively, found a
correlation coefficient of $\sim$0.85 in both cases (equal to the one
presented here). We cannot be certain of the nature of the
difference, but we speculate that the limitations of their explored
sample { (14 and 28 galaxies respectively, limited range of stellar masses, aperture
limitation of the MaNGA IFU data), could affect the strength of the
resolved rSFMS relation somehow. But we cannot be conclusive in this regards.}

In conclusion, we do not find a significant prevalence among the scaling relations
either at local or galaxy-wide scales, in contrast with previous studies. The results
for the different subsamples are contradictory, and the comparison between
the correlation coefficients and both the observed and intrinsic dispersion
of the relations do not provide with a consistent picture. In some cases the strongest
correlation does not correspond with the lowest intrinsic dispersion (e.g., CALIFA/rSFMS, EDGE/rSK relations), and in general the strongest correlation is not always the same of the different datasets (rSFMS for CALIFA and APEX, rSK for EDGE). Finally, in most of the cases there is
no significant difference between the observed standard deviations of the residuals and the
expected one due to error. The largest differences are found for the rSFMS, in the case of
EDGE and CALIFA ($\sim$0.05-0.06 dex), and for the rSK, in the case of APEX ($\sim$0.07 dex).
Even in these cases is difficult to justify their significance.


\subsection{Which parametrization best describes the \Ssfr?}
\label{sec:par}

In the previous sections we established that the three explored
parameters (\Sst, \Sgas and \Ssfr) correlate with one each other in a
similar way at both kpc-scales and averaged galaxy-wide. Following
\citet{lin19} and \citet{ellison21a}, we tried to establish a prevalence among the reported
relations. However, we did not find a clear one. The explanation
offered by the prevalence proposed by \citet{lin19} and \citet{ellison21a} is appealing.  To
consider the \Sgas\ as the main driver of the star-formation process
fits with our understanding of the physical process at low physical
scales, as first proposed in the seminal studies by \citet{schmidt59}
and \citet{kennicutt1998}, and explored recently using spatially resolved
CO observations \citep[e.g.][]{leroy13}. However, this view does not fit
with the picture emerging from the exploration of galaxy wide relations,
like the global (extensive) SFMS, in which the role of the depth of the
stellar potential seems to govern the SF \citep[e.g.][]{saintonge2011}. However,
the discovery of the resolved rSFMS relation \citep{sanchez13,wuyts13} required redefining
this interpretation for a local one, in which the SF depends on the local depth
of the gravitational potential traced by \Sst \citep[][]{ARAA}.

The quest for a more global star-formation law, that includes both
parameter (\Sgas and \Sst), or even additional ones that trace the
dynamics of galaxies and regions within galaxies, have led to
different functional forms. The seminal work of \citet{elme93},
\citet{wong02}, and \citet{blitz04,blitz06} established the importance
of the interstellar medium pressure, which depends on the
gravitational field and hence the stellar mass, on determining the
density of the star forming gas and the atomic to molecular
transition. \citet{shi11} and \citet{shi18} proposed a dependence
between \Ssfr and a combination of the other two parameters with the
form \Sgas$\times$\Sst$^{-\beta}$ (with $\beta$=0.5), following a
power-law as the one followed by the relations explored along this
study. This parametrization was motivated by the proposed effect of
the mid-plane pressure in self-regulating the star-formation process
\citep[e.g.][]{ostriker10}.  This parameterization was explored by
\citet{lin19}, finding that a value of $\beta$=0.3 slightly improves
the dispersion with respect to the one reported for the original rSK
one, for a slope of $\alpha$=1.38, in the overall power-law. Thus,
they proposed that the generalized star-forming law should exhibit the
form
$\Sigma_{\rm SFR} \propto \Sigma_{\rm mol}^{1.38} \Sigma_*^{-0.39}$.
However, we should note that they did attempt to attribute this
parametrization to the need for a secondary dependence, in an explicit
way, and indeed they concluded that the original rSK form is
sufficient to describe the observed distributions.

In an almost simultaneous study,
\citet{dey19} explored a more generalised relation to describe \Sst
using not only \Sgas and \Sst, but including a plethora of additional parameters, such as
both resolved and global properties of galaxies, and dynamical,
stellar population and ionized gas properties.  Their exploration was based on
a subset of 38 galaxies extracted from the current EDGE sample, although
using a preliminary extraction of both the CO and optically based dataproducts.
They used a Least
Absolute Shrinkage and Selection Operator to identify which of the explored
properties significantly influence the \Ssfr. Then, selecting those
that strongly predict the star-formation, they proposed a relation
based on eight parameters, with the functional form $\Sigma_{\rm SFR} \propto \Sigma_{\rm mol}^{0.4} \Sigma_*^{0.7} R^{0.6} \tau_*^{-0.6} \sigma_*^{0.2} Z_{\rm gas}^{-2.7}$, where $R$ is the galactocentric distance normalized by the effective
radius of the galaxy, $\tau_*$ is the mass-weighted stellar age,
$\sigma_*$ is the stellar velocity dispersion and $Z_{\rm gas}$ is the
solar-normalized gas oxygen abundance.

The first thing to notice is that both relations predict a totally
different dependence of \Ssfr on the \Sst. While the physically
motivated functional form explored by \citet{lin19} proposes a relation
in which \Sst presents a negative slope, on the contrary to \Sgas, the
empirical one proposed by \citet{dey19} uses a positive
slope for both parameters. In other words, the first relation follows
a trend consistent with the rSK, but not with the rSFMS, while the
second one follows a trend qualitatively consistent with both
relations.  

\begin{table*}
\caption[]{Results of the exploration of the distribution in the 3D space}
\label{tab:3D}
\begin{tabular}{llrrrrrr}
\hline
 Relation & \multicolumn{1}{c}{Dataset} & \multicolumn{1}{c}{$a$} & \multicolumn{1}{c}{$b$} & \multicolumn{1}{c}{$c$} &   \multicolumn{1}{c}{$d$} & \multicolumn{1}{c}{r$_c$} & \multicolumn{1}{c}{$\sigma$} \\
\hline
Plane & EDGE    &  0.439$\pm$0.005 & 0.566$\pm$0.005  &  -1  & 9.504$\pm$0.008 & 0.72 & 0.257\\
      & CALIFA  &  0.657$\pm$0.015 & 0.383$\pm$0.024  &  -1  & 9.887$\pm$0.049 & 0.85 & 0.246\\
      & APEX    &  0.466$\pm$0.021 & 0.323$\pm$0.019  &  -1  & 9.497$\pm$0.039 & 0.74 & 0.234\\
\hline
PCA   & EDGE    &  0.751$\pm$0.004 & 0.002$\pm$0.008  &  -0.662$\pm$0.004  &  6.810$\pm$0.031  &  0.64 &  0.307  \\
      & CALIFA &  0.734$\pm$0.021 &  0.076$\pm$0.043  &  -0.679$\pm$0.026  &  7.133$\pm$0.160  &  0.83 &  0.278  \\
      & APEX    &  0.697$\pm$0.112 & -0.195$\pm$0.390  &  -0.569$\pm$0.151  &  6.024$\pm$1.343  &  0.59 &  0.356  \\
\hline
Line  & EDGE    &  0.510$\pm$0.008 &  0.490$\pm$0.070  &  -1  &   9.555$\pm$0.180  & 0.77 & 0.232 \\
      & CALIFA &  0.505$\pm$0.075 &  0.475$\pm$0.105  &  -1  &   9.640$\pm$0.190  & 0.88 & 0.231 \\
      & APEX    &  0.370$\pm$0.105 &  0.380$\pm$0.135  &  -1  &   9.310$\pm$0.270  & 0.79 & 0.217 \\
  \hline
\end{tabular}

Best fitted parameters ($a$,$b$,$c$ and $d$), under the three
different assumptions included in this study (Plane, PCA or Line), and
the different explored dataset (EDGE, CALIFA and APEX), for the
functional form
$\Sigma_*^{a} \Sigma_{\rm mol}^{b} \Sigma_{\rm SFR}^{c} = 10^{d}$,
together with the correlation coefficient (r$_c$) and the standard
deviation of \Ssfr once the best fitted model is subtracted ($\sigma$).
For the "Line" we list parameters of the average between the rSK and rSFMS,
that would require also a rMGMS to define a line in the 3D space.

\end{table*}

If all the parameters adopted to build the independent parameter in both relations would be
totally independent, i.e., not presenting any relation among themselves, the
\Ssfr would follow a plane in the hyper-space of parameters.  In the
case of the functional form proposed by \citet{lin19} a plane in the
\Sst,\Sgas and \Ssfr three-dimension space (although we should clarify
that they never claim that the distribution follows a plane, and the dataset
shown in the 3D space in their Fig. 1 could be well described as a cylinder). On the other hand, if the
parameters entering in the equation are fully dependent to each
other, the \Ssfr should follow a line in the hyperspace (but not a
plane). Indeed, \citet{dey19} presented the existing relations between
the different additional parameters including in their functional
form. This result, and the existence of the rSFMS, rSK and rMGMS
relations themselves, explored in the previous sections, supports more
the second interpretation than the first one. In other words, the three parameters
\Sst,\Sgas and \Ssfr depend to each other. However, in this case, a different
sign in the slope for the \Sgas and \Sst in the proposed relations is
not clearly justified (since it naturally described a plane, not a
line).

We explore here three different possibilities to describe the relation
among the three considered parameters in order to understand which one
describes better the observed distribution. To do so we determine
which of them has the strongest correlation and the minimum
standard deviation of the residuals for the \Ssfr. We define those
three possibilities as:
\begin{itemize}
    \item { Plane:} We consider that \Ssfr is the truly dependent parameter,
    with \Sgas and \Sst being the independent ones. This way the distribution is described by 
    a plane with the functional form:
    \begin{equation}\label{eq:P}
\begin{matrix}
\Sigma_{\rm SFR} \propto \Sigma_*^{a} \Sigma_{\rm mol}^{b}
  \end{matrix}
\end{equation}  
    We
use the same least-square minimization adopted for the analysis of
the individual relations described in the previous sections to fit the data, with the
exception that we use the full set of data, not a binning scheme as the one
adopted before. { We should note that based on the results in the previous section,
  and recent results in the literature \citep{lin19,ellison20,sanchez20}, it is clear
  that choosing \Ssfr as the dependent parameter is somehow arbitrary. Furthermore, it
  is also clear that \Sgas and \Sst are not fully independent one each other. We adopted
  the current functional form following the literature in this topic, and in particular
the search for a generalized SF-law \citep{shi11,shi18,jkbb21a}}
   \item { PCA:} We consider that we do not know a priori
which is the independent parameter of the three. However, we assume that the distribution 
is well described by a plane in the 
three dimensional space, following the functional form:
\begin{equation}\label{eq:3D}
\begin{matrix}
\Sigma_*^{a} \Sigma_{\rm mol}^{b} \Sigma_{\rm SFR}^{c} = 10^{d} 
\end{matrix}
\end{equation}
Then, we perform a Principal Component Analysis (PCA hereafter) to determine the parameters
that describe better this plane, without assuming which is the dependent parameter (we adopted the PCA algorithm included in the {\tt scikit-learn} package implemented in {\tt python}). Note
that under this nomenclature the previously fitted functional form
would be the same, fixing $c=$-1. 
\item { Line:} Finally, we consider that the parameters follow a line in the 
three dimensional space (or 3D-line), instead of a plane. For a first approximation we derive this line 
as the intersection of the rMGMS ($\Sigma_*^{\alpha_{\rm rMGMS}} \Sigma_{\rm mol}^{-1} = 10^{-\beta_{\rm rMGMS}}$) with the plane resulting from the average of the sSK and rSFMS relations discussed before (Fig. \ref{fig:2D}, Tab. \ref{tab:2D}):
\begin{equation}\label{eq:avg}
\begin{matrix}
\Sigma_*^{0.5\alpha_{\rm rSFMS}} \Sigma_{\rm mol}^{0.5\alpha_{\rm rSK}} \Sigma_{\rm SFR}^{-1} = 10^{-(\beta_{\rm rSK}+\beta_{\rm rSFMS})}
\end{matrix} 
\end{equation}
\noindent This approach for defining a 3D-line allows us to compare the coefficients of Eq. \ref{eq:avg} with those provided by the two other methods (Eq. \ref{eq:P} and \ref{eq:3D}). However, it does not provide us with the
real functional form for the best three dimensional line describing the data, being just a first approximation. For completeness, we derive this line too by a simple linear regression to the data. Note that neither of these lines are the intersection of the previous fitted planes.
If that were the case we would not expect the line to improve the representation of the data.
\end{itemize}

\begin{table*}
\caption[]{Best fitted linear regression in the 3D space to a single line}
\label{tab:3D_line}
\begin{tabular}{lrrrrrrr}
\hline
\multicolumn{1}{c}{Dataset} & \multicolumn{1}{c}{$a_*$} & \multicolumn{1}{c}{$b_*$} & \multicolumn{1}{c}{$a_{\rm SFR}$} & \multicolumn{1}{c}{$b_{\rm SFR}$} &\multicolumn{1}{c}{ $a_{\rm mol}$} & \multicolumn{1}{c}{$b_{\rm mol}$}& \multicolumn{1}{c}{$\sigma$} \\
\hline
EDGE   & 1.740  $\pm$ 0.005 & 1.222 $\pm$ 0.011 &  0.683  $\pm$ 0.005 & 0.762  $\pm$ 0.013 & -8.322  $\pm$ 0.005 & 1.016 $\pm$ 0.014 &  0.185 \\
CALIFA &  1.862  $\pm$ 0.030 & 1.156 $\pm$ 0.067 & 0.615  $\pm$ 0.025 & 0.790 $\pm$ 0.045 & -8.445  $\pm$ 0.040 & 1.053 $\pm$ 0.045 &  0.248 \\
APEX &  2.033   $\pm$ 0.045 &  1.293 $\pm$ 0.073 & 0.691  $\pm$ 0.050 & 0.834  $\pm$ 0.081 & -8.332  $\pm$ 0.030 & 0.873 $\pm$ 0.087 &  0.241 \\
\hline
\end{tabular}

Best fitted parameters ($a_{par}$ and $b_{par}$), of the parametrization of the data in
the three-dimensional space by a line, following the functional form described in Eq. \ref{eq:3D_line}. The standard deviation of $\Delta$\Ssfr of the different datasets with respect to the best fitted line is also included ($\sigma$).
\end{table*}


This analysis and its results are illustrated in Figure \ref{fig:3D}, comprising
the three dimensional distribution of the considered parameters (\Ssfr, \Sst and \Sgas)
for the three different samples adopted in this article, together with the best
fitted models following the four approaches. These values, together with the correlation
coefficient and the dispersion once the best model is subtracted, for the first three adopted functional forms, are listed in Table \ref{tab:3D}.
As expected the first two approaches (Plane and PCA) do not provide linear distributions
along the 3D space. But instead both describe a plane, whose projection in the
three axes can not reproduce the three relations explored in the previous sections (rSFMS, rSK and
rMGMS). 

In the case of \Ssfr being a fully dependent parameter of the other two (Plane case),
the best fitted model produces qualitatively well the results by \citet{dey19}, with the two
powers of the \Sgas and \Sst parameters having the same sign (both positive). In the case of the CALIFA
and EDGE datasets the actual values of the slopes match within 0.1 dex with respect to the reported ones.
It is interesting to note the projection of the ``Plane'' relation passes through the distributions in the
\Ssfr-\Sst and \Ssfr-\Sgas planes, following the expected trends, but with considerably shallower slopes.
On the other hand, it cannot describe the distribution in the \Sgas-\Sst plane,
since it assumes that both parameters are independent. Curiously, the introduction of this functional form to describe the \Ssfr does not significantly improve the characterization of the data.
Both the correlation coefficients and the standard deviations of the residuals are the same or worse
than those reported for the rSFMS as listed in Tab. \ref{tab:2D}. In the case of the rSK, the correlation
coefficient of the ``Plane'' model is improved only for the APEX and CALIFA datasets, with a decrease
of the standard deviation only significant for the later one. For the APEX dataset, there is no
improvement for the rSK in any of the two parameters.

The PCA analysis tries to reproduce the plane in the space of
parameters using the linear combination of the parameters that retains most of
the variance of the distribution, thereby minimizing the
residuals with respect to the derived model. This analysis does not
require assuming which is the dependent parameter, as indicated
before. In this regard it is interesting to note that our current
analysis attributes the lowest variance to \Sgas, and it seems that
most of it is attributed to the combination  of \Ssfr and \Sst, i.e., the two
parameters involved in the rSFMS. Based on our results, this procedure
yields a plane in the 3D space, the projections of which reproduce
this relation better than any of the other two. Indeed, for the
\Sgas-\Sst plane, the projection seems to be orthogonal to the best
fitted rMGMS. The power law index (slope) of the relation attributed to \Sgas is
small, almost consistent  with zero. As in the case of
the Plane approach, the best fitted models do not provide  an
improvement of the characterization in terms of the standard
deviations of the residuals (Tab. \ref{tab:2D} vs. Tab. \ref{tab:3D}).
Actually, the residuals are larger than the ones produced  by the three
originally explored relations (rSFMS, rSK and rMGMS), for any of the
explored datasets. On the contrary, the correlation coefficients are
similar, just slightly worse (CALIFA and APEX), or slightly better
(EDGE) than the ones provided by the Plane-model. It is worth noting
that despite their physical motivation, neither the Plane nor the PCA procedures
reproduce the coefficients of the plane proposed by \citet{shi11} or \citet{lin19}.

\begin{figure}
 \minipage{0.48\textwidth}
 \includegraphics[width=\linewidth]{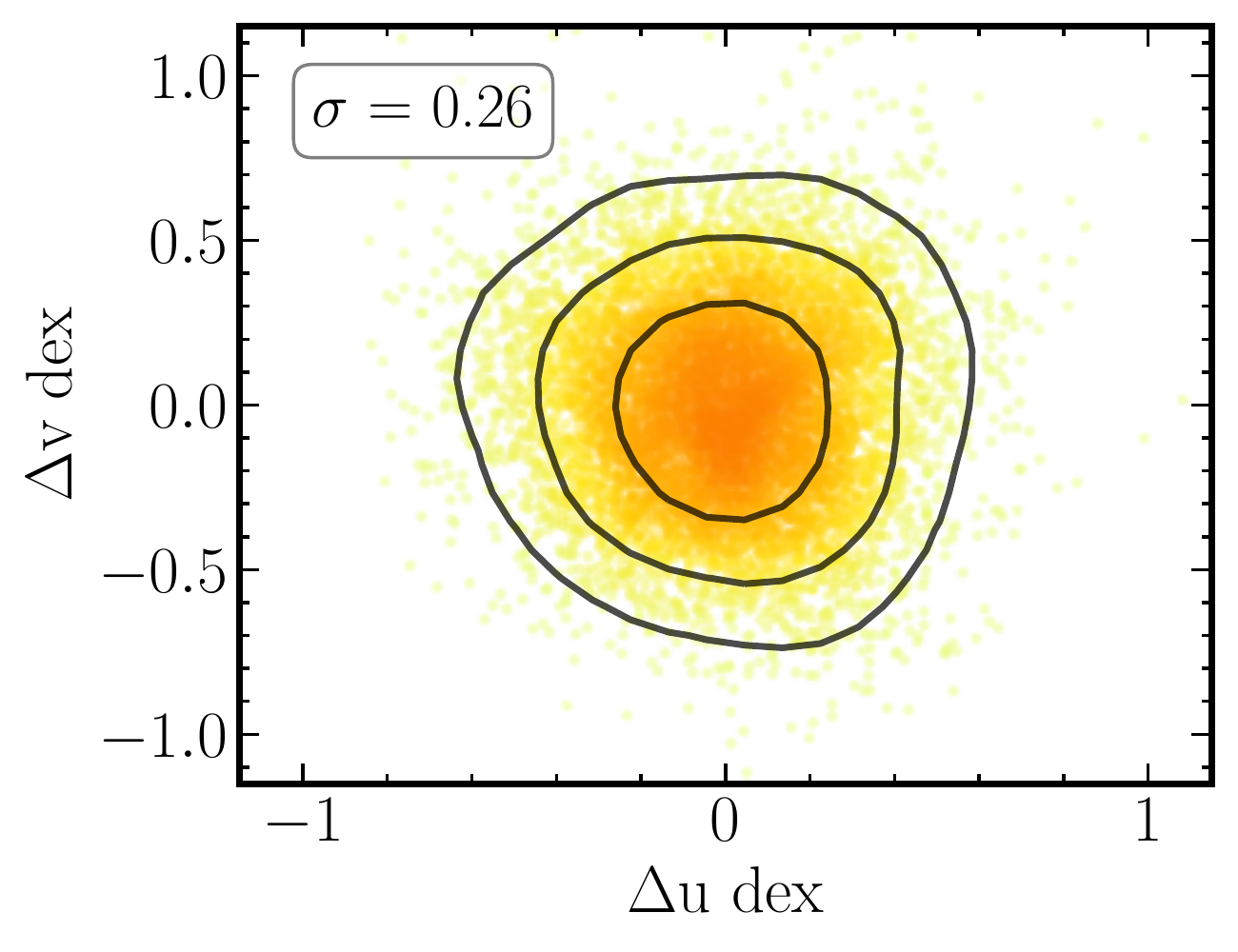}
 \endminipage
 \caption{Distribution of the residuals along the perpendicular plane to the best fitted
 3D-line representing the EDGE dataset (solid-line in Fig. \ref{eq:3D}, Tab. \ref{tab:3D_line}).
 Both axes ($u$,$v$) do not fully corresponds to any of the explored parameters, due to the projection, although $u$ is almost parallel to \Sgas and $v$ to
 \Ssfr (due to a pure mathematical artifact). Individual LoS are represented by solid circles, following the same scheme adopted in Fig. \ref{fig:2D}. The standard
deviation of the distribution in the y-axis is included in the corresponding inset. Contours in each panel corresponds to the area encircling 95\%, 80\% and 40\% of the datapoints (black solid lines).}
 \label{fig:d_3d}
\end{figure}


The 3D line resulting from the average of the rSFMS and rSK relations (Eq. \ref{eq:avg}), together with the rMGMS reported in the
previous section (values in Tab. \ref{tab:2D}),
shows the best characterization of the data. The standard deviation of
the residuals is clearly the lowest of the three explored approaches,
for the three datasets. It presents a lower standard deviation that the original rSFMS and
rSK relations too. It also exhibits the strongest correlation coefficients,
improving upon the previous parametrizations in  all datasets.  By construction, it reproduces
the two relations from which it is created. We note that if, instead of combining the rSFMS and rSK to reproduce the
distribution in the 3D space we adopt any other combination of relations, we would derive very similar results. This is a consequence of the fact that the three parameters (\Sst, \Ssfr and \Sgas) follow a line in the 3D space. For instance, it
is easy to demonstrate that if the data fulfill simultaneously the rSK and rSFMS relations, then \Sgas and \Sst cannot be independent parameters. By just solving the equation:
\begin{equation}\label{eq:rMGMS}
\begin{matrix}
\beta_{\rm SFMS}+\alpha_{\rm SFMS} {\rm log}(\eSst) = {\rm log}(\eSsfr) =  \beta_{\rm SK}+\alpha_{\rm SK} {\rm log}(\eSgas) 
\end{matrix}
\end{equation}
we would derive a rMGMS relation, the coefficients of which are fully consistent with the ones listed in Tab. \ref{tab:2D}. This is apparent in Fig. \ref{fig:3D}, where
we show that this inferred rMGMS relation (represented as dotted-dashed line in Fig. \ref{fig:3D}) is in excellent agreement with the best fitted one (represented
as a solid-line in the same figure). It is important to note that we were unable to attain such a result when  using  the Plane or PCA parametrizations of the relations.

\begin{figure*}
 \minipage{0.99\textwidth}
 \includegraphics[width=\linewidth]{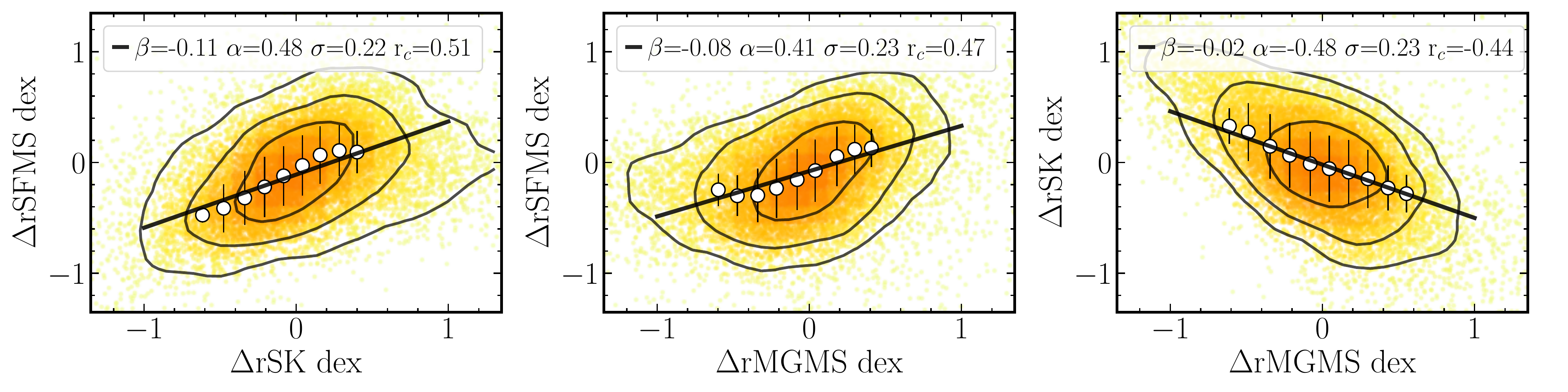}
 \endminipage
 \caption{Distribution of the residuals of the three relations explored along this study one versus another:
   (i) $\Delta$rSFMS vs $\Delta$rSK (left panel); (ii) $\Delta$rSFMS vs. $\Delta$rMGMS (central panel); and (iii)
   $\Delta$rSK vs. $\Delta$rMGMS (right panel). Individual LoS (solid circles, color coded by the density of
   points), mean values in bins of 0.15 dex of the parameter in the X-axis (white circles), and the best fitted linear
   regression (solid line) are represented in each panel following the same scheme adopted in Fig. \ref{fig:2D}.
   The parameters of the best fitted linear regression, following Eq. \ref{eq:log}, together with the standard
   deviation of the residuals and the correlation coefficients between the datasets shown in each panel are
   included in the corresponding inset. Contours in each panel corresponds to the area encircling 95\%, 80\% and 40\%
   of the datapoints (black solid lines).}
 \label{fig:Delta_Delta}
\end{figure*}

Finally, we find the best parametrization of this 3D line  using
a simple linear regression of the dataset, corresponding to the black-solid line shown in each panel
of Fig. \ref{fig:3D}. This line has the functional form:
\begin{equation}
\label{eq:3D_line}
\begin{pmatrix}
{\rm log} \eSst  \\
{\rm log} \eSsfr \\
{\rm log} \eSgas\\
\end{pmatrix} = 
\begin{pmatrix}
a_*\\
a_{\rm SFR}\\
a_{\rm mol}\\
\end{pmatrix}
+
\begin{pmatrix}
b_*\\
b_{\rm SFR}\\
b_{\rm mol}\\
\end{pmatrix}
\times t
\end{equation}
\noindent where $t$ is the independent parameter tracing the linear distribution, and parameters $a$ and $b$ are the coefficients describing the 3D line for each explored physical property. This parametrization can be written in a functional form more similar to Eq. \ref{eq:rMGMS}, by solving \Ssfr:

\begin{equation}\label{eq:solv}
\begin{split}
\frac{{\rm log}(\eSst) - a_{\rm *}}{b_{\rm *}} = \frac{{\rm log}(\eSsfr) - a_{\rm SFR}}{b_{\rm SFR}} = \frac{{\rm log}(\eSgas) - a_{\rm mol}}{b_{\rm mol}} \\
b_{\rm SFR}\frac{{\rm log}(\eSst) - a_{\rm *}}{b_{\rm *}} + a_{\rm SFR} = {\rm log}(\eSsfr) = b_{\rm SFR}\frac{{\rm log}(\eSgas) - a_{\rm mol}}{b_{\rm mol}}  + a_{\rm SFR}
\end{split}
\end{equation}

\noindent This way it is possible to compare the coefficients of the parametrization to those listed in Tab. \ref{tab:2D} and \ref{tab:3D}.
Table
\ref{tab:3D_line} lists the estimated values for the coefficients and the standard deviation of the residual with respect to the best fitted line. For the three datasets we derive very similar coefficients, in particular for the slope ($b$) of the linear relation. The steeper slope corresponds to the one for the \Sst relation (being slightly higher than one), and the shallower one for the \Ssfr (being slightly lower than one). The residuals for \Ssfr around the best fitted 3D line are very similar among the different datasets, being of the order of those reported for the average line (Table \ref{tab:3D}).

As a final check, we present in Figure \ref{fig:d_3d} the distribution of residuals across the plane perpendicular to this best fitted 3D-line for the EDGE dataset (the axis are labeled $u$ and $v$, representing the directions perpendicular to the line in Eq. \ref{eq:3D_line}). If the distribution along the 3D-space is sufficiently well represented by a 3D-line, then these residuals should not present any elongation. In contrast, if a plane describes better the observed distribution, those residuals should present a detectable elongation along the direction defined by the plane itself. As apparent in the figure the residuals present no elongation along a preferred direction, with a negligible correlation coefficient. 
We note that in principle the axes represented in this figure do not necessarily correspond to any of the three physical quantities explored here. Nonetheless, it turns out that $v$ is almost parallel to \Ssfr with a scaling factor $\sim0.762$ (the slope of the 3D-line for this parameter, Tab. \ref{tab:3D_line}).
We also note that the  standard deviation in Fig. \ref{fig:d_3d} fully matches the one enlisted in Tab. \ref{tab:3D_line} for \Ssfr, once the scaling factor is taken into account. Similar results
are found for the APEX and CALIFA datasets. 

In summary, this final test reinforces our results showing that the distribution in the 3D space is best  represented by a line, from which the previous relations are pure projections. Therefore, there is
no preference among the rSFMS, rSK and rMGMS relations in their description of the observed distributions.
We will discuss the implications later in Sec. \ref{sec:dis}.


\subsection{Exploring possible secondary relations}
\label{sec:sec}


In the two previous sections we have characterized the paired
relations between the \Ssfr, \Sst and \Sgas (i.e., the rSFMS, rMGMS
and rSK relations), showing that they follow a log-log relation
in the space defined by them that is well
characterized by a 3D-line. This analysis shows that
the distribution is not well represented by a plane or a surface. This,
in essence, limits the possibility of a third parameter that
significantly contributes to the dispersion in the analyzed relations,
at least a parameter among the ones explored here or well-correlated with them. In other words, if the
relation followed by the three parameters is a line and not a plane,
the residual in any of the explored relations (e.g., the rSFMS) should
not depend significantly on any parameter that includes the third
explored parameter ({ this is, \Sgas, in the case case that the explored relation is rSFMS}). However, previous
studies have indicated that some of these relations show explicit
dependencies with a third parameter. For instance, \citet{rosa16} and
\citet{mariana19} show that the the rSFMS segregates by morphology,
with SF regions in later-type galaxies having slightly larger
\Ssfr for a fixed \Sst than earlier types. { Since morphology is
  directly connected with star-formation stage, $f_{gas}$ and SFE \citep[e.g.][]{saint16,calette18,sanchez18}, then any morphological difference may be interpreted as a difference induced by a third additional parameter.
Indeed, some authors \citep[e.g.][]{ellison20} have reported
that the observed scatter in the rSFMS depends both on the
star-formation efficiency (SFE) and the molecular gas fraction ($f_{mol}$), with
a prevalence for the first parameter. Others explorations of the integrated SFMS (
for only SFGs) suggest that the scatter depends more on the gas fraction rather than on the SFE
\citep[e.g.][]{saint16,sanchez18,colombo20}. In some cases, the}  differences in the relation
found galaxy by galaxy have led to some authors to claim that indeed
there is no general rSFMS relation \citep[e.g.][]{vulcani19}, 
which is totally contrary to most previous results \citep[as reviewed
by][]{ARAA}. { Finally, the} search for a third
parameter to reduce the scatter of the observed relations has led to
modifications of the rSK law, which includes \Sst
\citep[known as the extended Schmidt-law, e.g.][]{shi11} and other resolved and integrated parameters
\citep[e.g.][]{dey19}, as indicated in the previous section.

The main hypotheses behind these different explorations are that (i)
indeed the scatter in the relations depends on a third parameter, (ii)
the residual of the considered relation correlates with either a
parameter (or property), a combination of parameters, or the residuals
of another relation, and (iii) that this new correlation is physically driven,
significant, decreasing the scatter around the original relation and
altering significantly its shape. For instance, Table 2 of
\citet{mariana19} shows that the shape of the rSFMS (its intercept
and slope) for SF regions is different for different morphological types. In
the same way, \citet{ellison20}, shown in their Fig. 4, that the
residual of the rSFMS strongly correlates with that of the rSK (a parameter
that tightly correlates with the SFE if the slope of rSK is near to one). However, none of them shows a really clear and
significant decrease of the original scatter when these possible
secondary relations are introduced. On the contrary, \citet{dey19} and
\citet{lin19} show a decrease of the scatter of the original
relation (rSK and rSFMS, respectively), when additional parameters are
included in the parametrization of the relations.

\begin{figure*}
 \minipage{0.99\textwidth}
 \includegraphics[width=\linewidth]{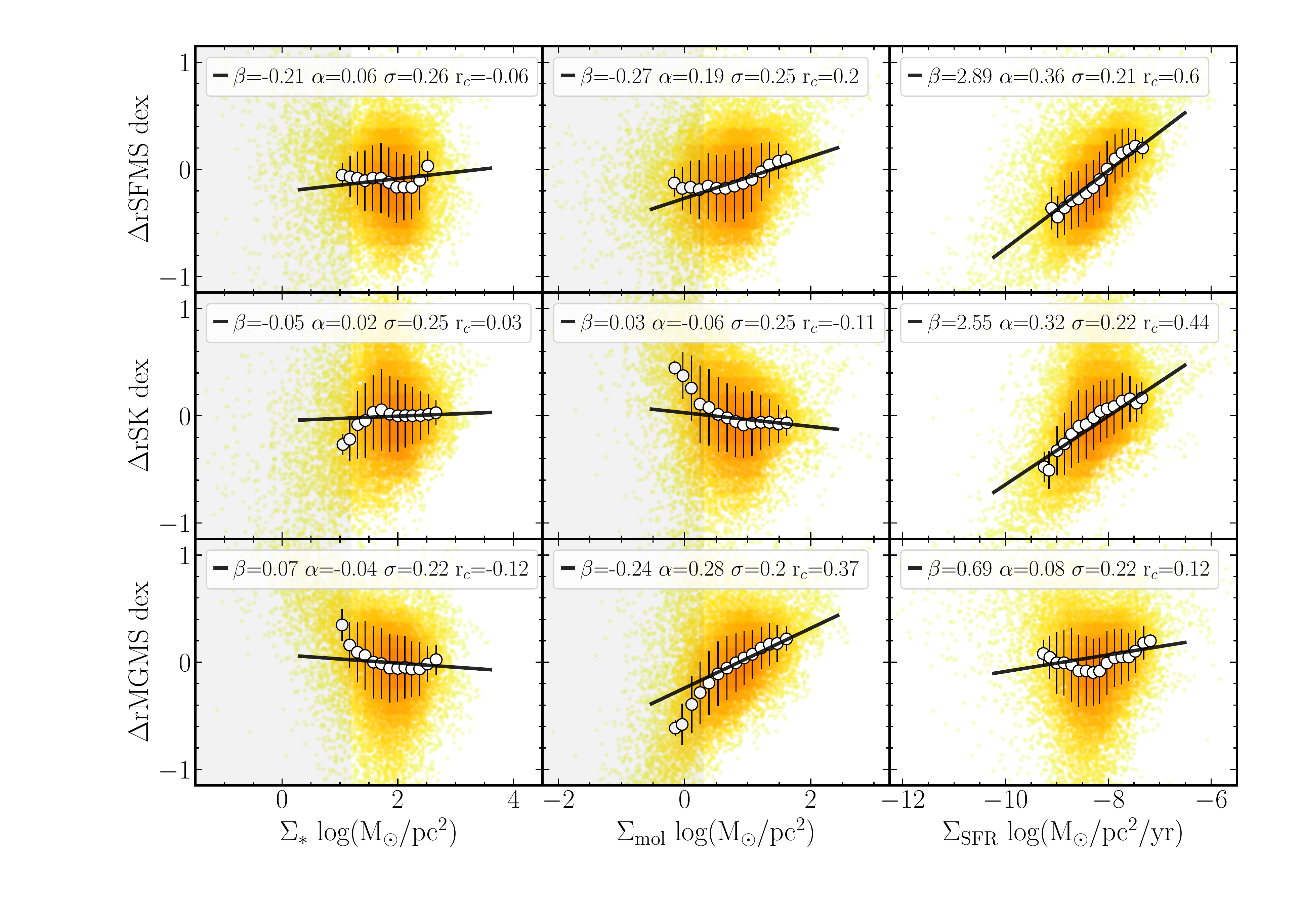}
 \endminipage
 \caption{Residuals of the rSFMS (top panels), rSK (middle panels), and rMGMS (bottom panels) along
   the \Sst (left panels), \Sgas (middle panels) and \Ssfr (right panels) for the EDGE dataset.
   Individual LoS (solid circles, color coded by the density of
   points), mean values in bins of 0.15 dex of the parameter in the X-axis (white circles), and the best fitted linear
   regression (solid line) are represented in each panel following the same scheme adopted in Fig. \ref{fig:2D}.
   The parameters of the best fitted linear regression, following Eq. \ref{eq:log}, together with the standard
   deviation of the residuals and the correlation coefficients between the datasets shown in each panel are
   included in the corresponding inset. }
 \label{fig:Delta}
\end{figure*}

Based on the previous section results, we would expect 
little or no dependence either between  the residuals of the
explored relations or between those residuals and a third parameter
not considered in the relations. To explore this possibility in the
most ample and agnostic way we first explore the residuals of the
three relations one versus each other
\citep[following][]{ellison20} that is, the relations between $\Delta$rSFMS, $\Delta$rSK and $\Delta$rMGMS where residual is defined as the 
removal of the ``best fit'' relation from the measurement of the dependent parameter (for example, removing the rSFMS from the measured value of \Ssfr for a given \Sst).
Figure \ref{fig:Delta_Delta} shows the
results of this exploration. Contrary to what would be naively expected
from the results presented in the previous sections there is a clear
correlation between the three residuals. The strongest one is the one
between $\Delta$rSFMS and $\Delta$rSK, that it is just slightly less
strong and shows the same slope as the one presented by
\citet{ellison20}, in their Fig. 4.  It is followed by the relation
between $\Delta$rSFMS and $\Delta$rMGMS, and finally by the one
between $\Delta$rSK and $\Delta$rMGMS. The correlation coefficients 
in the three cases are larger than 0.4. In the case of the rSFMS there is a small but appreciable
decrease of the scatter with respect to the original relation for both the
tentative secondary relation with the $\Delta$rSK and $\Delta$rMGMS.
If these secondary relations were physically motivated, our conclusions
would be similar to the ones presented by \citet{ellison20}, suggesting
that the scatter of the rSFMS is driven mostly by the SFE ($\sim$$\Delta$rSK, in particular
when $\alpha=$1). Furthermore, we could conclude that the scatter in the rSK is partially
due to gas fraction (f$_{\rm gas}$$\approx$$\Delta$rMGMS).

\begin{figure*}
 \minipage{0.99\textwidth}
 \includegraphics[width=\linewidth]{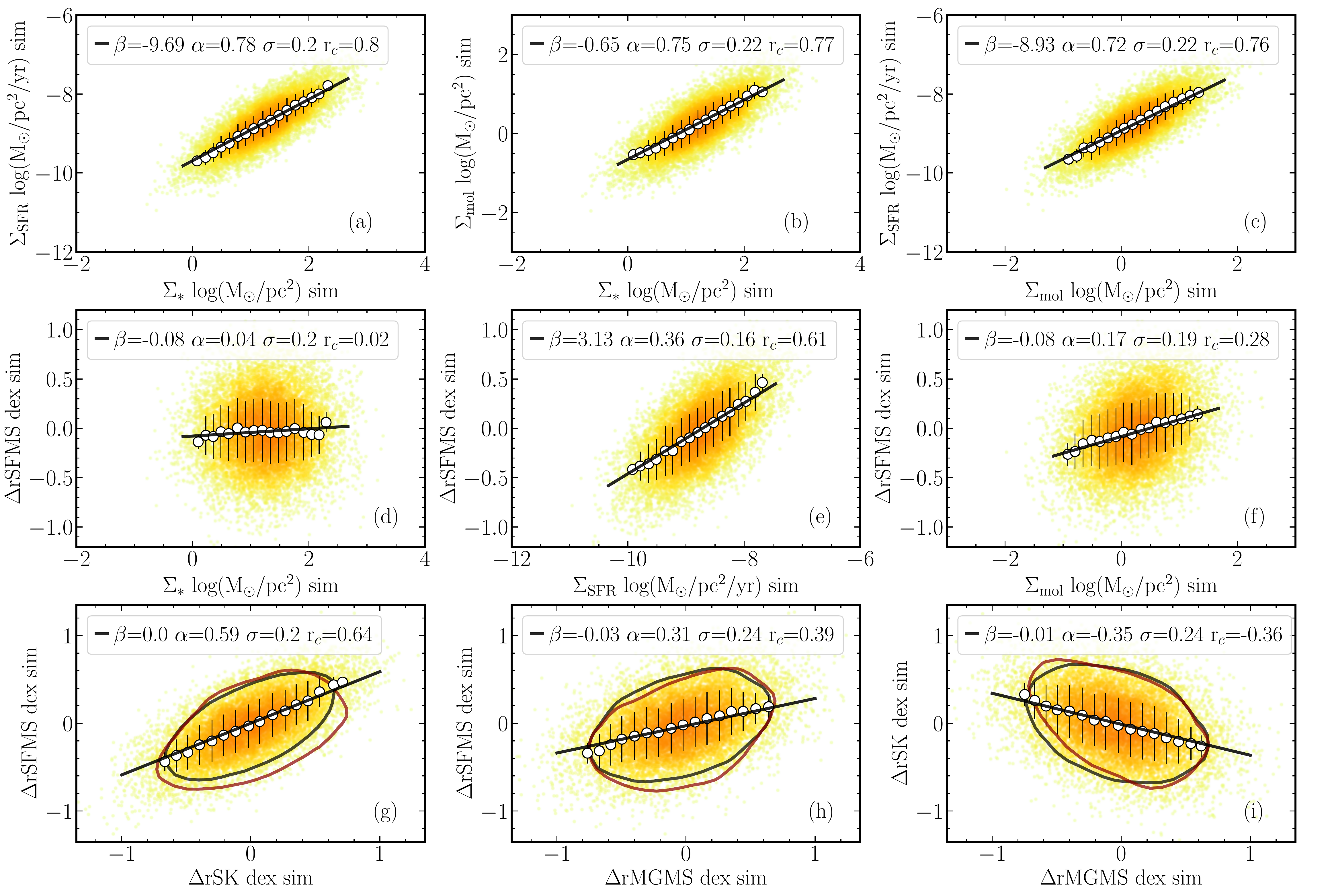}
 \endminipage
 \caption{Results of the exploration of a simple simulated dataset comprising 10,000 spatially resolved SF regions
   following a linear rSFMS ($\alpha=$1, $\beta=$-10) and a linear rMGMS ($\alpha=$1, $\beta=$-1),
   with \Sst distributed randomly around an average value of 10$^{1.25}$ M$_\odot/$pc$^2$ following a standard
   normal distribution with a standard deviation of of $\sigma=$0.675 dex. The typical errors corresponding to the observed
   dataset are simulated by adding random noise to each
   of the considered parameters, \Sst, \Ssfr and \Sgas, following a simple Gaussian
   distribution with a standard deviation of 0.20 dex, 0.20 dex and 0.22 dex, respectively. Top panels
   show the simulated rSFMS (a) and rMGMS (b), together with the recovered rSK (c). Central panels
   show the residual of the rSFMS, after subtracting the best fitted linear regression, against \Sst (d),
   \Ssfr (e) and \Sgas (f). Symbols, insets and labels are similar to the ones adopted in Fig. \ref{fig:Delta}.
   Finally, bottom panels show the distribution of the residuals of the three relations shown in the
   top panels versus each other: $\Delta$rSFMS vs. $\Delta$rSK (g), $\Delta$rSFMS vs. $\Delta$MGMS (h)
   and $\Delta$rSK vs. $\Delta$MGMS. Thus, the bottom panels correspond to the same distributions shown in Fig. \ref{fig:Delta_Delta},
   but for the simulated distributions. To highlight the similarities between both distributions, we include
   the density contour of the EDGE dataset encircling the 80\% of the points shown in the quoted figure  (dark-red solid lines), together with the same density contour for the simulated data (black solid line). }
 \label{fig:sim}
\end{figure*}

It this were correct, however, these results would be totally inconsistent with those shown
in the previous section. To understand in more detail the origin of
the observed trends between the residuals, we explore their dependence
on the original explored parameters: \Sst, \Ssfr and \Sgas.  Figure
\ref{fig:Delta} shows the distributions of these residuals against each other. For
completeness, we present the comparison of each residual against each
parameter, including the independent parameter of the respective
relation. Indeed, having removed the primary dependence, it is
expected that the residual of each relation  does not
depend on the corresponding independent parameter  (for example, $\Delta$rSFMS should not depend on \Sst).  However, as a sanity check, it is important to
demonstrate so. If a dependence were found if would mean that the fitted model is not a good
representation of the data, retaining some dependence with the independent parameter (e.g., \Sst).
In this case any secondary relation could be just an
effect of an incomplete removal of the primary one.
For example, a residual dependence of $\Delta$rSFMS with \Sst would result in a artificial relations with both \Ssfr and \Sgas as a consequence of the rSFMS and rMGMS relations. 
Fig. \ref{fig:Delta}
demonstrates that our fitted models removed  most of the variation of
their independent parameter. For all three residuals
($\Delta$rSFMS, $\Delta$rSK, and
$\Delta$rMGMS), the residual correlations with their original independent parameter have a very low correlation coefficient ($r_c$ in the $0.03-0.09$ range), and they present a negligible slope
($\alpha\approx-0.08$ to $0.07$) with no significant improvement over the
original standard deviation (see Tab. \ref{tab:2D}). Curiously, the
contrary happens with the corresponding dependent parameter in each
relation (\Ssfr in $\Delta$rSFMS, \Ssfr in $\Delta$rSK, and \Sgas in $\Delta$rMGMS).  In all three cases there
is always a weak/moderate correlation
($r_c\sim0.37-0.59$), a non negligible slope
($\alpha\sim$0.3), and a small but measurable decrease in the
standard deviation of the residuals that can be as large as $\sim$20\%.

This indicates that despite the fact that the linear relations shown in Fig. \ref{fig:2D} and Tab. \ref{tab:2D}
describe the distribution of observed parameters, they cannot describe the distribution of the
errors. These errors induce non negligible residuals, once  the
best fitted relation is subtracted, in both the independent and the dependent parameter  and those
residuals correlate with each other. The reason is that any deviation from the main relation due to an error induces a residual that depends on this error modulo the slope of the relation, as we will show in the next section. This residual will correlate with the dependent parameter too,
since the best fitted relation for the observed parameters minimizes the correlation
of the residual of the dependent parameter with the independent one, but it does not minimize the correlation
of that residual with the dependent parameter. This reasoning  suggest that the described
secondary relations shown in Fig. \ref{fig:Delta_Delta} and \ref{fig:Delta} may not be physical in nature, which we will explore in detail in the next section.



%

\subsection{Are secondary relations physical?}
\label{sec:real}

At first glance, the results from the two previous sections seem
  contradictory since our best fitted models removed most of
the dependence with respect to the independent parameters, 
but their residuals retain a significant dependence on the dependent parameter of the original relation.  
{ In order to prove if the reported relations are physical or if they are
  induced artificially by the individual errors we have performed a set of simple simulations
  in which we assume that : (1) the three parameters present a set of linear relations between
  them; (2) there is no real secondary relation. Thus, the residual of any relation between
  two of the parameters does not depends on the third one; and finally (3) the individual errors are
  Poissonian and fully independent between the three parameters (i.e., there is no co-variance).
  Based on these simulations we try to understand how reliable are the recovered parameters from our fitting scheme and if secondary relations may be induced by errors.}

In a more mathematical form, let us
assume that three datasets ($x$,$y$ and $z$) are described by a set of
linear relations:

\begin{equation}
 \begin{split} 
 YX : y = a_{xy} x + b_{xy}  \\ YZ : y = a_{zy} z +  b_{zy}  \\ ZX : z = a_{xz} x + b_{xz}  \\
 \end{split}
 \label{eq:rel}
\end{equation}

\noindent if coefficents $a_{ij}$ and $b_{ij}$ (where $i$ and $j$ take the values $x$,$y$ or $z$, for $i\ne j$)
are perfectly well determined, then the residuals of the three
relations ($\Delta YX$, $\Delta YZ$ ,$\Delta ZX$) should be zero by
construction. However, in the presence of noise, this is not
true. Let's assume that the three datasets are measured in a
fully indepedent way. That is, the errors in the measurements are not
correlated in any way. In this case, the observed datasets would not
be $x$, $y$ and $z$, but:

\begin{equation}
 \begin{split} 
   x_{obs} = x + e_{x}, \\
   y_{obs} = y + e_{y}, \\
   z_{obs} = z + e_{z}, \\
 \end{split}
 \label{eq:err}
\end{equation}

\noindent where $e_{i}$ is the error of variable $i$ (an $e_j$ is
totally independent of $e_i$, or any $i \neq j$). In this case the
residuals of the relations would be:
\begin{equation}
 \begin{split} 
   \Delta YX_{obs} = e_{y} - a_{xy} e_{x}, \\
   \Delta YZ_{obs} = e_{y} - a_{zy} e_{y}, \\
   \Delta ZX_{obs} = e_{z} - a_{xz} e_{x}, \\
 \end{split}
 \label{eq:d_err}
\end{equation}

\noindent for a perfect determination of the coefficients $a_{ij}$
and $b_{ij}$. 
If not,  
a term $e_{b,ij}$+$e_{a,ij} i$ need be subtracted from 
each equation, to
account for the error in the derivation of the coefficients 
(where $e_b$ corresponds to the error in the intercept and $e_a$ to the error in the slope). So
far we consider negligible those errors for clarity. In this case, it is possible
to find correlations (or their absence) between the residuals and the
relations and the dependent (or independent) parameters and among the
residuals themselves under certain circumstances. It is out of the
scope of the current exploration to cover all the possible
combinations, which are regulated mostly by (i) the range of values
covered by each parameter (characterized by its standard
deviation around the mean value $\sigma_i$), relative to the error of the
considered parameter ($e_i$); (ii) the comparison between the errors
among the different parameters; and (iii) the slope of the three
relations among them ($a_{ij}$). If $\sigma_i$ is not orders of
magnitude larger than $e_i$, and $a_{ij}$ is near one in the three
cases, it is expected both a correlation between the residuals and the
dependent parameter (since they both includes the term $e_y$ or
$e_z$), and among the residuals themselves (with $\Delta YX$ and
$\Delta YZ$ positively correlated due to the presence of $e_y$ in both
of them, and $\Delta YZ$ and $\Delta ZX$ anti-correlating, since the
first depends on $-a_{zy}\,e_z$, and the second in
$e_z$).  The strength of each correlation would depend on the actual
values of $\sigma_i$, $e_i$ and
$a_{ij}$, and the best simplest way to assess them is through a simulation. A
comparison between these {\it induced} correlations and the observed
relations would establish the physical origin (or not) of the later
ones.

For the current exploration, we identify the three
parameters ($x$, $y$, and
$z$) with \Sst, \Ssfr, \Sgas, in
logarithmic scale.  We adopt a typical
$\sigma_x=$0.5 dex for the current dataset, and a typical error of
$e_i=$0.26 dex, generating a dataset of 10,000 simulated values
(per parameter) \footnote{for different datasets we refer the reader
  to the following script
  \url{https://github.com/sfsanchez72/error_sim/blob/master/error_simulation.ipynb}}.
The simulated values follow the relations described in Eq. \ref{eq:rel},
assuming as intercepts
($b_{rSFMS,YX}$=-10.0 dex
$b_{rMGMS,ZX}$=-1.0 dex), and values of \Sst around
10$^{1.25}$ (i.e., $x\sim$1.25$\pm$0.4 dex). These values are adopted
to match the observations, although they are unimportant to assess the
strength of the error-induced correlations.

Figure \ref{fig:sim} presents the main results of this simulation. The
upper panels show the simulated distributions between the three
considered parameters. 
The parameters exhibit correlations of similar or slightly higher strength than the observed ones, as our simulated error distribution is very simple and
does not include systematic effects,
detection limits, correlated noise or non-homogeneous sampling of the
distribution as would be expected in real data. 
For each distribution we repeated the analysis we performed on the observed data in Sec. \ref{sec:sec} using the same code. This analysis provides
the  power-law relations between the three parameters, with
best fitted values very close to those used to create the simulated data.  The
standard deviation of the residuals once the best fit
relations are subtracted are similar to the ones reported for the observed
distributions ($\sigma\sim$0.2 dex). As with the real data
(Fig. \ref{fig:Delta}) we explore the possible dependence of the
residuals of each simulated relation on each considered parameter.
We present in Fig. \ref{fig:sim} (central row panels) the results for the
simulated $\Delta$rSFMS, as similar distributions are
observed in the other two residuals. Similar to the case of the real
data, the simulated residuals show no correlation with the independent
parameter of the considered relation (\Sst, r$_c=$0.03, $\alpha$=-0.01), a weak positive trend with the parameter not involved
in the relation (\Sgas , r$_c$=0.24, $\alpha$=0.13), and a clear
positive correlation with the dependent parameter of the considered
relation (the parameter in the ordinate axis, i.e., \Ssfr r$_c$=0.59, $\alpha$=0.33).  The agreement with the
observed distributions is very good (compare Fig. \ref{fig:Delta} and
\ref{fig:sim}). Furthermore, the simulation exhibits clear
correlations between the residuals of the three relations:
Fig. \ref{fig:sim}, bottom panel, shows the same distributions shown
in Fig. \ref{fig:Delta_Delta}, but for the simulated dataset. For
comparison purposes we include the contour corresponding to the 80\%
encircled density of the observed distribution together with that of the
simulated ones. The agreement between the shape of the distributions,
and the trend and strength of the correlations is remarkable good
considering the simplistic nature of the simulation.

Most likely, including the possible correlations between the errors
derived from the same dataset (like \Sst and \Ssfr), and further fine
tuning to match them with the observed distribution in a
better way would produce even more similar results. Nonetheless,
these simulations are sufficient to demonstrate that the observed secondary
trends between the residuals of the explored relations (rSFMS, rSK and rMGMS)
and between these residuals and the involved parameters (\Sst, \Ssfr and \Sgas)
are a pure consequence of the noise in the observed datasets. In other
words, these relations are most likely a pure mathematical artifact and have no physical origin.

\section{Discussion and Conclusions}
\label{sec:dis}

In this study we have explored the local/resolved and global
intensive relations between the surface densities of the stellar mass,
star-formation rate and molecular gas mass, using the combined dataset
provided by the CALIFA survey and the spatially resolved and aperture
integrated CO observations provided by the CARMA and APEX antenna
(within the framework of the EDGE collaboration). This exploration
comprises the largest dataset of galaxies with spatially resolved
molecular gas combined with IFS optical data so far (more than 100
galaxies, and more than 10,000 independent line of sights), and one of
the largest dataset of galaxies with aperture matched extimates of the three considered parameters (more than 400
galaxies).

\subsection{The connection betweem local and global relations}

Using this unique dataset, we have characterized the resolved and
global relations between the three described parameters (\Sst, \Ssfr
and \Sgas), for the star-forming regions (galaxies), i.e., the rSFMS,
rSK and rMGMS relations (and their corresponding unresolved counterparts). In the case of the resolved properties we
confirm that the three relations are well characterized by a single
power-law, with a slope near to one, and a
small scatter ($\sigma\sim$0.2 dex), as previously described by
other authors \citep{sanchez13,wuyts13,cano16,lin19,ellison21a}. We show 
that the intensive global relations
follow similar trends as the resolved/local ones, in the three cases, as
suggested by previous authors \citep[for each
relation][]{pan18,mariana19,ARAA}. Indeed, when the distributions of
the star-forming regions (and galaxies) are compared in the same
parameter space (\Ssfr-\Sst, \Ssfr-\Sst and \Sgas-\Sst) they are
statistically indistinguishable. In other words, they are the same
relations. In a recent exploration of this topic, \citet{sanchez20}
demonstrated that indeed if the resolved regions of galaxies follow a
local/resolved relation, the existence of a global one is a pure
mathematical derivation from the local one. Using simulations this was
already demonstrated for other local relations, like the resolved
Mass-Metallicity relation \citep[rMZR or \Sst-O/H relation,
e.g.][]{rosales12}. However, this is the first time, to our knowledge,
that this is demonstrated for the three explored relations, using such an
ample dataset, and using { both direct and indirect} estimates of the molecular gas via
CO observations { and dust-to-gas calibrations}.

The main implication of the match between the resolved and global
intensive relations, and the prevalence of the first ones
\citep{sanchez20}, is that the main physical processes that regulate
star-formation operate at the spatial scales where the resolved ones
are found (i.e., at kpc-scales), and not by other processes that affect
galaxies as a whole. This does not imply that global processes does not affect
the star-formation process \citep[as discussed in][]{ARAA}. Galaxy interactions among themselves or with the environment, galactic-wide outflows, and the ignition of an active nuclei
are some of the global processes that indeed affect the star-formation galaxy wide  \citep[][]{hopkins+2010,GASP,sanchez18,bluck19}. However, they do not
have a direct effect galaxy wide. They affect the whole galaxy through 
the physical processes that regulate star-formation at the explored resolved scales.

On the other hand, it is known that star-formation
happens at much smaller scales than the ones explored here, i.e., the size of molecular clouds
(from a few pc to some hundred pc). Newly born stars are created when
these clouds fragment and collapse gravitationally. This simple
argument was used by \citet{schmidt59} to propose a relation between
the stellar mass formed in a certain time and the amount of gas in the
volume at which the stars are formed. From these arguments, assuming
that all new stars are formed in the characteristic free-fall time of
the gas \citet{kennicutt1998} estimated that the relation between both
quantities should follow a power-law of $\sim$1.5. However, this
relation cannot be the same as the one we observe at kpc-scales or galaxy
wide. The observed relation does not match the formed stellar mass in
a free-fall time-scale with the pre-existing gas mass. It relates
the average SFR with the average gas mass on much larger scales.
The same holds for the rSFMS and rMGMS relations. Indeed, the simple
free-fall scheme does not even work at the scale of the molecular clouds too: for an already
collapsed molecular cloud in which gas mass has been transformed to stars, the
SFR may be the previous gas mass (if all gas is transformed to stars)
divided by a characteristic time-scale in which SF happens (with the
free-fall time being a good proxy). The emerging relation would
connect the previously existing molecular gas mass with the newly
created stellar mass (a rMGMS-like relation), or the already happened
SFR with both quantities (previously existing or already existing gas
and stellar masses, i.e., a rSFMS-like relation). But the existence of
those physical relations does not guarantee that the average SFR
within a certain time scale and a much larger spatial scale would be related
to the current reservoir of molecular gas or the previously
formed stellar mass. If that was the case all molecular gas should
inevitably collapse at a rate dictated by the free-fall time scale,
and all SF should happen in a single burst that consumes all the
existing gas.

However, we do not observe this in galaxies or regions
within galaxies. Not all the molecular gas is collapsing. Other
physical conditions are required for that to happen \citep[e.g.][]{elme97}. 
The hypothesis explored here
is that SFR is somehow regulated by the previously formed stellar
content, that prevents or allows the collapse of new molecular clouds. 
Then, once collapsed, they would individually and locally follow the relation proposed by \citet{kennicutt1998} (or a similar one). Therefore, this relation
is diluted at kilo-parsec scales by the averaging between regions
actually forming and regions still not forming (or with stars already
formed) producing a slope in all the observed relations that is
shallower ($\sim 1$) than that predicted by the simple free-fall time
scheme (1.5). Furthermore, this relation is only valid when averaged
over much larger times scales than that of the SF process itself (for
instance, the free-fall time scale). This regulation is most probably provided by the SF feedback (including supernovae and stellar wind), as proposed by many different studies
\citep[][]{silk97,ostriker10,hopkins2013,kruij19}. In this case, the typical time
scale should correspond to the one required by the characteristic velocity of
a stellar wind ($\sim$10-30 km/s) to propagate through the considered
physical scales ($\sim$1kpc). This corresponds to $\sim$30-100 Myr.
Numerical simulations indeed show that these are the typical spatial- and
time-scales at which a well-defined quasi-steady-state exists in which the energy
injected by SFR feedback pressurizes on average the ISM \citep[][]{kim17,kim18,seme17,Orr18}, which we speculate leads to the observed relations.
If this is the regulation mechanism, then the considered relations should
break at spatial scales on which the observed time-scale for the SFR
(measured by H$\alpha$ would correspond to $\sim$4-10 Myr) is not
large enough for the wind to propagate through the considered spatial
regime. This would correspond to $\sim$40-100 pc, a regime at which
the described relations has not been explored well. Indeed, the same
numerical simulations indicate that at these smaller spatial- and time-scales the SFR
and the properties of the ISM exhibit strong local variations. Observationally it is seen a break-down of the rSK relation at happens somewhere between 80-100 pc \citep[][]{ono10,ver10,williams18} and 500 pc \citep[][]{Kennicutt07,blanc09}, which could indicate that additional processes affect the regulation process in different galaxies or environments.


As we show in Sec. \ref{sec:ana}, the three involved quantities are
tightly correlated which each other, describing not only a set of
relations between any pair of them (Fig. \ref{fig:2D}), but also they are
tightly correlated in the 3D space formed by themselves
(Fig. \ref{fig:3D}), following { a power law}. Being unclear which of them is the primary
independent parameter, a reasonable conclusion is that a physical
quantity that parametrizes the effects of the stellar feedback
\citep[e.g.][]{silk97,lilly13}, would be the hidden parameter that
explains the three relations simultaneously. Current simulations
\citep[e.g.][]{ostriker10} and observations \citep[e.g.][]{leroy08,sun20} suggest
that the mid-plane pressure, generated mostly by stellar winds and modulated by
the gravitational potential and the gas content, could be the hidden
parameter behind the explored relations \citep[e.g.][]{jkbb21a,jkbb21b}.

\subsection{Are secondary relations needed?}

The second most important result from our exploration is that the
dispersion about the best derived relations that characterize the
resolved relations (rSFMS, rSK and rMGMS) is totally driven by the
errors in the individual parameters (Sec. \ref{sec:real}). Despite the
fact that there are appreciable and (in principle) statistically
significant secondary correlations between the residuals of these
relations and the explored parameters and among the residuals
themselves (Sec. \ref{sec:ana}), our results indicate that they are
not physically driven secondary relations, but a pure mathematical
artifact of the covariance between axes driven by individual errors or errors. This result has profound
implications, since recent explorations have interpreted these
secondary relations as physically drivers for the dispersion
\citep{mariana19,lin19,ellison20,ellison21a,colombo20}, motivated either by the
variations in the relative amount of gas ($f_{gas}$, related to
$\Delta$rMGMS), changes in the SFE (related to $\Delta$rSK), or even
other additional properties (like the morphology). { \citet{ellison21a} reported 
variations up to an order of magnitude in all the three relations between
different galaxies, what may dominate the scatter in the average relations.
In summary, they proposed a non-universality of these kpc-scale relations, in
the same line of the results by \citet{vulcani19}.}

On the contrary, our results suggest that so far an additional
physical driver may not be required to explain the dispersion,
which is fully attributable to the errors in the individual parameters.  This does not mean that
certain regions (and galaxies) do not depart from these relations because of
physical reasons. Galaxies/regions with extreme SFE (e.g., Starburst,
ULIRGS), would be clearly above the rSFMS and rSK relations
\citep[e.g.][]{ellison20b}, and on the contrary, galaxies/regions with
low $f_{gas}$ would be well below the rSFMS and the rMGMS relations
\citep[e.g.][]{saint16,calette18,sanchez18,sanchez20}. Consequently,
earlier-type galaxies (and regions within them), with lower $f_{gas}$
would depart from the corresponding relations. However, this corresponds to
regions/galaxies that depart from the observed trends, and not to
galaxies well represented by those trends. This was recently discussed in
\citet{colombo20}, where there is a clear different trend with SFE and
$f_{gas}$ for galaxies within the SFMS and SK relations (that we now
understand are driven by errors) and for galaxies departing from
those relations (where the effect is most probably physical).


\subsection{Main caveats to our results}

Despite the fact that our sample is larger in number of galaxies, and it comprises a much
larger number of independent resolved LoS than most of the
previously studied ones, the dataset is not free of biases and potential problems.
First, our physical spatial resolution ($\sim$ 1.3 kpc) is not good enough
to distinguish individual star-forming regions (even giant \HII\ regions),
and certainly it is not good enough to perform an optimal separation
between different ionizing sources \citep[e.g., the diffuse][]{lacerda18,espi20}.
Therefore, a fraction of the reported dispersion could be due to
mixing between regions with different ionizations, and different SF regions. However,
following our primary result, the scaling relations seem to have the same characteristic shape
(and dispersion) at any physical scale within galaxies (larger than the smaller sampled
scale). Therefore, the relatively coarse spatial resolution should not have
a significant/strong effect above 500 pc. Obviously, exploring these relations at higher/better
physical resolution would improve our understanding of them \citep[e.g.][]{schruba11,kruij14}, in particular to
determine at which scale the relations are expected to break (demonstrating
the hypothesis that they are statistical relations{ , discussed before}), and limit the effect
of mixing on the accuracy of the estimated parameters which produces
an increase of the observational uncertainties. 

Second, the EDGE sub-sample, which we use to explore the resolved
relations, is still biased towards late-type galaxies in comparison
with the CALIFA and APEX sub-samples. Furthermore, it has a
relatively shallow depth. Therefore, although it represents a clear
improvement over other samples and datasets it may need to be further improved
to properly span the parameter space, for example to break the scaling relations
by morphology, or to
explore in detail the loci of retired regions in the considered
diagrams (Fig. \ref{fig:2D}). Because of this we cannot be totally
conclusive.  
Note, however, that many of the studies regarding the physical drivers of the dispersion
around the explored relations base their analyzes on datasets with even stronger sample biases
(see the comparison in Fig. \ref{fig:sample}). In any case, 
observations of the CALIFA sample with ALMA would be key to improve our understanding
of these relations: the CALIFA IFS data are the only ones that span a very large and uniform sample of galaxies with coverage out to 2.5 R$_e$ that is well-matched to the field-of-view of ALMA. 

\subsection{Summary of the conclusions}

Based on the analysis presented in this study we conclude that:
\begin{itemize}
\item The intensive form of global and local relations among \Ssfr, \Sst and \Sgas are
  essentially the same relations, following a set of power-law with the same slopes and zero-points, when
  expressed in the same units.
\item These relations hold at very different scales, from galaxy wide, to scales involving
  a fraction of the galaxies ($\sim$1/3 of the galaxies), to kiloparsec scales.
\item They are expected to break at lower scales (100-500 pc), highlighting their
  statistical nature, what indicate that some-kind of self-regulation process that happens
  at kiloparsec scales is behind those relations.
\item The three explored parameters (\Ssfr, \Sst and \Sgas) follow a
  line (or cylinder) in the three dimensional space, with the relations between each pair of them
  (known as rSFMS, rSK and rMGMS) just projections of that relation.
\item The scatter around this relation (and its projection) is fully
  dominated by individual errors, without the need of a secondary
  relation with one of the parameters when the relation between the
  other two is removed.
\end{itemize}


Finally, we highlight the importance of understanding the effects of
axis covariance and uncertainties in the exploration of possible
relations of the residuals of primary relations. An approach such as
the one outlined in this study need be applied in future explorations,
and to revisit claims of secondary relations with the required detail.
In future explorations with the current dataset we will try to
understand the underlying nature of the physical parameters driving
the described relations \citep{jkbb21b}, and the causes of the halting
of the star-formation in retired regions in galaxies.



\section*{Acknowledgements}

We acknowledge the comments and suggestions by the anonymous referee that
has helped to improve this manuscript.

Authors thanks S. Ellison and L. Lin for their comments and
suggestions that have significantly improved the manuscript.

SFS and J.B-B are grateful for the support of a CONACYT grant CB-285080 and FC-2016-01-1916, and funding from the PAPIIT-DGAPA-IN100519 (UNAM) project. J.B-B acknowledges support from the grant IA-100420 (DGAPA-PAPIIT ,UNAM) and funding from the CONACYT grant CF19-39578. DC acknowledges support from \emph{Deut\-sche For\-schungs\-ge\-mein\-schaft, DFG\/} project number SFB956A. T.W., Y.C., and Y.L. acknowledge support from the NSF through grant AST-1616199. 
A.B., R.L. and S.V. acknowledge support from the NSF through gran NSF AST-1615960.

Support for CARMA construction was derived from the
Gordon and Betty Moore Foundation, the Eileen and Kenneth Norris
Foundation, the Caltech Associates, the states of California,
Illinois, and Maryland, and the NSF. Funding for CARMA development and
operations were supported by NSF and the CARMA partner
universities. 

This study uses data provided by the Calar Alto Legacy
Integral Field Area (CALIFA) survey (http://califa.caha.es/). Based on
observations collected at the Centro Astron\'omico Hispano Alem\'an
(CAHA) at Calar Alto, operated jointly by the Max-Planck-Institut
f\"ur Astronomie and the Instituto de Astrof\'isica de Andaluc\'ia
(CSIC). 

This research made use of
Astropy,\footnote{http://www.astropy.org} a community-developed core
Python package for Astronomy \citep{astropy:2013, astropy:2018}.

\section*{Data Availability}

The data used in this article consist of the publicly available DR3
of the CALIFA survey as well as extended surveys such as PISCO. CALIFA
DR3 data cubes can be accessed at
\url{https://califaserv.caha.es/CALIFA_WEB/public_html/?q=content/califa-3rd-data-release},
with direct access to the V500 version used in this article available
at \url{ftp://ftp.caha.es/CALIFA/reduced/V500/reduced_v2.2/}.


The EDGE CO datacubes are publicly available at \url{https://www.astro.umd.edu/EDGE/}.

The APEX data will be delivered in a forthcoming article (Colombo et
al. in prep.). A preliminary description of the data is included in \citep{colombo20}


\bibliographystyle{mnras}
\bibliography{my_bib} 

\end{document}